\documentclass[twocolumn,nofootinbib,prd,aps,amsmath,amssymb,superscriptaddress]{revtex4-1}
\usepackage{mathtools}
\usepackage{amsthm}
\usepackage{bm}
\usepackage{graphicx,epsfig}
\usepackage{ulem}
\usepackage{xfrac}
\usepackage{color}
\usepackage{cancel}
\usepackage{bigints}

\newtheorem{theorem}{Theorem}
\newtheorem{corollary}{Corollary}
\newtheorem{lemma}{Lemma}
\newtheorem{proposition}{Proposition} 
\newtheorem*{remark}{Remark}
\usepackage{soul}
\usepackage{hyperref}
{
 \definecolor{BLACK}{gray}{0}
 \definecolor{WHITE}{gray}{1}
 \definecolor{RED}{rgb}{1,0,0}
 \definecolor{GREEN}{rgb}{0,1,0}
\definecolor{dgreen}{rgb}{.1,.6,.1}
\definecolor{BLUE}{rgb}{0,0,1}
 \definecolor{CYAN}{cmyk}{1,0,0,0}
 \definecolor{MAGENTA}{cmyk}{0,1,0,0}
 \definecolor{YELLOW}{cmyk}{0,0,1,0}
 \definecolor{aw}{rgb}{0.2,0.5,0.75}
  }
  \hypersetup{
    colorlinks,%
    citecolor=blue,%
    filecolor=blue,%
    linkcolor=magenta,%
    urlcolor=blue
}

\usepackage{mathtools}

\newcommand{\mathsout}[1]
{\bgroup\mathchoice
  {\sbox0{$\displaystyle{#1}$}%
    \usebox0\hspace{-\wd0}%
    \rule[0.5\ht0-0.5\dp0-.5pt]{\wd0}{1pt}}%
  {\sbox0{$\textstyle{#1}$}%
    \usebox0\hspace{-\wd0}%
    \rule[0.5\ht0-0.5\dp0-.5pt]{\wd0}{1pt}}%
  {\sbox0{$\scriptstyle{#1}$}%
    \usebox0\hspace{-\wd0}%
    \rule[0.5\ht0-0.5\dp0-.5pt]{\wd0}{1pt}}%
  {\sbox0{$\scriptscriptstyle{#1}$}%
    \usebox0\hspace{-\wd0}%
    \rule[0.5\ht0-0.5\dp0-.5pt]{\wd0}{1pt}}%
\egroup}

\newcommand{\ba}{\begin{eqnarray}}
\newcommand{\ea}{\end{eqnarray}}

\newcommand{\bse}{\begin{subequations}}
\newcommand{\ese}{\end{subequations}}

\newcommand{\W}{{\cal {W}}}
\newcommand{\bbq}{\begin{quote}}
\newcommand{\eeq}{\end{quote}}

\newcommand{\CR}{{\cal{R}}}
\newcommand{\T}{{\cal{T}}}

\newcommand{\G}{{\cal{G}}}
\newcommand{\A}{{\cal{A}}}
\newcommand{\F}{{\cal{F}}}
\newcommand{\s}{{\cal{S}}}
\newcommand{\dd}{{\hbox{d}}}

\def\t#1#2#3{#1^{\if#2- \else #2 \fi}_{\if#2- \else \:\: \fi #3}}

\def\tdot#1#2#3{\dot{#1}^{\if#2- \else #2 \fi}_{\if#2- \else \:\: \fi #3}}
\def\tddot#1#2#3{\ddot{#1}^{\if#2- \else #2 \fi}_{\if#2- \else \:\: \fi #3}}
\def\ttilde#1#2#3{\tilde{#1}^{\if#2- \else #2 \fi}_{\if#2- \else \:\: \fi #3}}
\def\a#1#2{\eta^{#1}_{\:\:#2}}
\def\ba#1{\boldsymbol{\eta}^{#1}}

\def\adot#1#2{\dot{\eta}^{#1}_{\:\:#2}}
\def\addot#1#2{\ddot{\eta}^{#1}_{\:\:#2}}

\font\bigastfont=cmr10 scaled \magstep 2
\def\bdot{\hbox{\bigastfont .}}
\newcommand{\dotaverage}[1]{\left\langle #1 \right\rangle^{\bdot}_\cD}

\newcommand{\dotaverageH}[1]{\left\langle #1 \right\rangle^{\bdot}_{\cD_H}}

\newcommand{\cD}{{\cal D}}
\newcommand{\cB}{{\cal B}}
\newcommand{\average}[1]{\left\langle #1 \right\rangle_\cD}
\newcommand{\averageH}[1]{\left\langle #1 \right\rangle_{\cD_H}}
\newcommand{\averageL}[1]{\left\langle #1 \right\rangle_{\cB}}

\newcommand{\CQ}{{\cal Q}}
\newcommand{\inI}{{\rm I}}
\newcommand{\inII}{{\rm II}}
\newcommand{\inIII}{{\rm III}}

\newcommand{\J}{\mathfrak{J}}

\newcommand{\hs}{\hat{\s}}

\newcommand{\z}{\xi}
\newcommand{\y}{\varkappa}
\newcommand{\x}{\varsigma}
\newcommand{\rv}{\boldsymbol{\xi}}

\begin{document}
\title{Lagrangian theory of structure formation in relativistic cosmology. \\VI. Comparison with Szekeres exact solutions}
\author{Ismael Delgado Gaspar}
\email{ismidelgado@astro.unam.mx}
\affiliation{Instituto de Astronom\'\i a, Universidad Nacional Aut\'onoma de M\'exico, AP 70-264, Ciudad de M\'exico, 04510, M\'exico}
\author{Thomas Buchert}
\email{buchert@ens-lyon.fr}
\affiliation{Univ Lyon, Ens de Lyon, Univ Lyon1, CNRS, Centre de Recherche Astrophysique de Lyon UMR5574, F-69007, Lyon, France}
\begin{abstract}
We examine the relation between the Szekeres models and relativistic Lagrangian perturbation schemes, in particular the Relativistic Zel'dovich Approximation (RZA). We show that the second class of the Szekeres solutions is exactly contained within the RZA when the latter is restricted to an irrotational dust source with a flow-orthogonal foliation of spacetime. In such a case, the solution is governed by the first principal scalar invariant of the deformation field, proving a direct connection with a class of Newtonian three-dimensional solutions without symmetry.  
For the second class, a necessary and sufficient condition for the vanishing of cosmological backreaction on a scale of homogeneity is expressed through integral constraints. Domains with no backreaction can be smoothly matched, forming a lattice model, where exact deviations  average out at a given scale of homogeneity, and the homogeneous and isotropic  background is recovered as an average property of the model.  
Although the connection with the first class of Szekeres solutions is not straightforward, this class allows for the interpretation in terms of a spatial superposition of nonintersecting fluid lines, where each world line evolves independently and under the RZA model equations, but with different associated ``local backgrounds''.
This points to the possibility of generalizing the Lagrangian perturbation schemes to structure formation models on evolving backgrounds, including global cosmological backreaction.
\end{abstract}
\pacs{04.20.-q, 04.20.Jb, 98.80.-k, 04.25.Nx}
%
\maketitle

\section{Introduction}

In the Lagrangian formulation, the Newtonian theory is paraphrased within the general relativity framework through  coframe fields, which constitute a formal generalization of the deformation gradient in Newtonian cosmology~\cite{Kasai1995,Sabino1996,rza1,rza2,rza3,rza4,rza5}.
Then, following Zel'dovich's extrapolation idea, other variables are functionally expressed in terms of this deformation field (for details on the transformation of the $3+1$ Einstein equations to a system for spatial coframes, see Ref.~\cite{rza1}, for the average properties of the first-order scheme, see Ref.~\cite{rza2}, for the $n^{th}$-order Lagrangian perturbation and solution schemes, see Ref.~\cite{rza3}, and for generalizations including the tensor perturbations, Ref.~\cite{rza4}, and including pressure through a change of foliation, Ref. \cite{rza5}).

The Relativistic Zel'dovich Approximation (RZA) forms an extrapolation of the first-order scheme, as defined in \cite{rza1}. It holds, by construction, nonlinearities encoded in the functional dependence of variables on the coframe fields, which accounts for the correct causal structure  and measurement of distances. However, despite approximate assumptions, it is remarkable that RZA contains as particular cases a subclass of the Szekeres solutions, furnishing the most general exact solutions applicable to cosmology. This was first noticed by M. Kasai in his seminal work~\cite{Kasai1995}, and we now revisit and prove it using the definition of RZA provided in the series of papers following \cite{rza1}.  A further remarkable property of RZA is that its spatial average also contains classes of averaged exact solutions, {\it e.g.} the spatially flat Lema\^\i tre-Tolman Bondi (LTB) solution, further discussed below. 

Szekeres models form a class of exact solutions to Einstein's equations, which, in general, present no symmetries~\cite{Sz75,Bonnor1977}                  
 (but quasisymmetries). Their field source is an irrotational but inhomogeneous dust fluid;  the spacetime is compatible with the inclusion of a cosmological constant 
 ($\Lambda$)~\cite{kras1,kras2,ellis2012relativistic,szafron1977inhomogeneous}. 
The solution is classified into two classes, depending on whether the metric function $\beta_{,\z}$ in their general line element (see equation~\eqref{Eq:GeneralSzeMetric} below) is different or equal to zero~\cite{kras1,kras2,BKHC2009}.  While the first class (class I: $\beta_{,\z}\neq0$) has been successfully used in cosmology and astrophysics~\cite{hellaby1996null,HellabyKrasi2002,Bolejko2006Struformation,bolejko2007evolution,Hellaby:2007hq,IshakNwankwo2008,Bolejko:2009GERG,BolCelerier2010,KrasBol2011,Nwankwo_2011,BolSuss2011,IshakPeel2012LargeScale,SussBol2012,WaltersHellaby2012,Buckley2013,IshakPRL2013,Vrba:2014,Koksbang2015,Koksbang2015II,sussman2015multiple,sussman2016coarse,GIraHellaby2017,Hellaby2017,gaspar2018black}, the second one (class II: $\beta_{,\z}=0$) has received much less attention. Among the exceptions we find~\cite{ishak2012growth,IshakPeel2012LargeScale,SzeLamddanot0Meures}.

In this paper we aim at addressing the question of which subclass of the Szekeres solutions is contained within RZA. We will call this subclass the `exact body of RZA'. The importance of this result is twofold: first, the Szekeres solutions can be used as a reference to test the accuracy of the functional evaluation of RZA that goes beyond a mere perturbative evaluation, and, second, RZA provides a guide for reinterpreting the Szekeres arbitrary functions in terms of  generalized Newtonian quantities. To examine the connection between the solutions, we analyze each class of the Szekeres solutions separately. While the relation of RZA to class I seems intricate and demands more future work, class II is exactly contained within RZA, and corresponds to a class of three-dimensional, locally one-dimensional Newtonian solutions without symmetries investigated in~\cite{Buchert1989AA} (see also the same solution class without a background \cite{buchertgoetz}, and the same class that includes the cosmological constant in the background \cite{bildhaueretal}).

The motivation underlying this work goes beyond these technical clarifications. The relativistic generalization of the Newtonian Lagrangian perturbation theory, with its first-order member RZA, reveals the powerful property that its average, 
\cite{rza2}, contains the spatially averaged exact spherically symmetric LTB solution \cite[Sect. 7.2]{BuchertFocus}, a property that is unexpected since the local model contains a class of plane-symmetric solutions and is expected to perform best for highly anisotropic collapse.  However, this remark holds true for flat LTB solutions only and as such it corresponds to the situation of Newton's iron sphere theorem. Thus, RZA appears to be a restricted answer to a full relativistic generalization, and we aim at understanding the class I Szekeres solutions as providing hints toward such a generalization.

The plan of this article is as follows. We begin by presenting the most fundamental properties of the Relativistic Zel'dovich Approximation and Szekeres models in Sections~\ref{Sec:RZAintro} and~\ref{Sec:SzekeresModelsintro}, respectively. Section~\ref{Recipe} provides the general steps to reformulate the Szekeres solutions in the language of relativistic Lagrangian perturbations. This reformulation is specialized to be compatible with RZA in Section~\ref{Sec:ClassIIRZA}, with the result that the whole class II is exactly contained in RZA. Within this section, we study the conditions under which deviations from an FLRW (Friedmann-Lema\^\i tre-Robertson-Walker) background solution average out on some scale of homogeneity, Subsection~\ref{SubSecNoteAdm}, while in Subsection~\ref{SubSec:CosmoLattice} we present a lattice model made up of consecutive cells with null backreaction on a particular homogeneity scale, smoothly matched across suitable surfaces. In Subsection~\ref{SubSec:RZAandCPT} we show that the RZA functionals reproduce the correct Szekeres quantities and examine the correspondence between class II and a class of three-dimensional Newtonian solutions without symmetry. In Section~\ref{Sec:CommentsClassI} we discuss the relation of class I solutions to RZA, reinterpret the dynamics as a set of independent world lines, and show that each one follows the RZA model equations. Our results are summarized and discussed in Section~\ref{Sec:DiscFinalRemarks}. 

The main text is complemented with nine appendices, providing the necessary background material to keep the paper as self-contained as possible:
Appendix~\ref{SecApp:RelBetweenPars} contains a detailed discussion about the relation between the Szekeres-Szafron and Goode-Wainwright parametrizations, which incidentally proves the compatibility of the Goode-Wainwright formulation (of both classes I and II) with the presence of a cosmological constant. 
As a reference to the case with $\Lambda=0$, in Appendix~\ref{App:ParametricSols}, we show the Goode-Wainwright parametric solutions of the Szekeres field equations. 
Appendix~\ref{App:FLRW-Cartesians} presents the transformations to Cartesian coordinates of some subcases of the Szekeres solutions.
The spatially averaged equations for the volume-expansion and volume-acceleration are shown in Appendix~\ref{App:BackAveEqns}.
Appendix~\ref{App:ProofTheor2} contains the formal proof of the Lemma~\ref{Lem:lemma2} enunciated in Section~\ref{SubSecNoteAdm}, while the proofs of Lemmata~\ref{Lem:ConsRho-3R} and~\ref{Lem:VanishingIntBeta} are provided in Appendix~\ref{AppSec:ProofOfLemmata3-4}.  
Supplementary calculations, based on the noncommutativity of averaging and evolution as well as cosmological backreaction, aim at a better understanding of the class I solutions and are presented in Appendix~\ref{AppSec:ComRulesCIassI}. 
For better readability, we have reserved for Appendix~\ref{SecApp:FuncEvaluationRZA} the functional evaluation of the relevant dynamical fields. 
Finally, in Appendix~\ref{App:LTBSubcase}, we provide the formal relation between LTB models and RZA, which supports the discussion conducted in Section~\ref{SubSec:AdmissibDataClassI}.

\section{The relativistic Lagrangian formulation}\label{Sec:RZAintro}

In this section, we summarize the most important results about RZA that are relevant to the subject of the paper. Therefore, we will limit our exposition to the case of an irrotational dust source with a flow-orthogonal foliation of the spacetime (compatible with the restrictions obeyed by the Szekeres solutions). For more details and generalizations including tensor perturbations (giving place to gravitational waves) or pressure gradients see~\cite{rza1,rza2,rza3,rza4,rza5} and \cite{BMRFoliations}. 

For comoving and synchronous observers, the metric takes the form:
\begin{equation}\label{gmetricI}
^{(4)} \mathbf{g} = -\mathbf{d}t  \otimes \mathbf{d}t + ^{(3)}\!\mathbf{g} \quad
\hbox{with} \quad
^{(3)}\mathbf{g}=g_{ij} \mathbf{d}X^i \otimes \mathbf{d}X^j \ ,
\end{equation}
where $X^i$ are Gaussian normal (Lagrangian) coordinates.\footnote{Indices $i, j, k, \cdots$ denote coordinate indices, while indices $a,b,c \cdots$ are introduced as counters of components, e.g. of vectors or differential forms. In this paper we use units where the gravitational constant and the speed of light are set to $G=c=1$.}
Following \cite{rza2,rza3,rza4,rza5}, the spatial metric is decomposed in terms of the coframes as follows:
\begin{equation}\label{Eq:LineElemRZAGab}
^{(3)}\mathbf{g}=G_{ab} \bm{\eta}^a \otimes \bm{\eta}^b \quad; \quad g_{ij}=G_{ab}\eta^a_{~i} \eta^b_{~j} \ ,
\end{equation}
where the coframes  are split into a trivial set and deviations thereof:
\begin{equation}\label{Eq:CoframesGab}
\bm{\eta}^a=\eta^a_{~i} \mathbf{d}X^i =a(t) \left(\delta^{a}_{~i}+P^a_{~i}\right) \mathbf{d}X^i \ .
\end{equation}
The initial metric coefficients are encoded in Gram's matrix $G_{ab}$:
\begin{equation}
G_{ab}(\mathbf{X})\delta^a_{\ i}\delta^b_{\ j} = G_{ij}(\mathbf{X})\equiv g_{ij}(t_{i},\mathbf{X}) \ .
\end{equation}

Through the $3+1$ formalism with a flow-orthogonal foliation of spacetime, Einstein's equations are transformed into a  system of $9+4$ evolution equations (the $4$ constraint equations of general relativity are transformed to evolution equations in the Lagrangian framework) for the 9 coframe coefficient functions~\cite{rza1,rza4}.
The complete system of equations reads:
\bse\label{Eq:CoframeRelRicci1}
\begin{align}
\label{form_symcoeff}&\hspace{7em} G_{ab} \,\dot{\eta}^a_{[i} \eta^b_{\ j]} = 0 \ ;
 \\
\label{form_eomcoeff}&\frac{1}{2 J} \epsilon_{abc} \epsilon^{ikl}  \left( \dot{\eta}^a_{\ j} \eta^b_{\ k} \eta^c_{\ l} \right)^{\bdot} = -\CR^i_{\ j} + \left( 4 \pi \varrho + \Lambda \right) \delta^i_{\ j} \ ;
 \\
\label{form_hamiltoncoeff}&\frac{1}{2J}\epsilon_{abc} \epsilon^{mjk} \dot{\eta}^a_{\ m} \dot{\eta}^b_{\ j} \eta^c_{\ k} = - \frac{\CR}{2}+ \left( 8\pi \varrho +  \Lambda  \right) \ ;
 \\
\label{form_momcoeff}&\left(\tfrac{1}{J}\epsilon_{abc} \epsilon^{ikl} \dot{\eta}^a_{\ j} \eta^b_{\ k} \eta^c_{\ l} \right)_{||i} = \left(\tfrac{1}{J}\epsilon_{abc} \epsilon^{ikl} \dot{\eta}^a_{\ i} \eta^b_{\ k} \eta^c_{\ l} \right)_{|j} \ ,
\end{align}
\ese
where the overdot stands for the covariant (here simply the partial) time-derivative; the single and the double vertical slash denote the ordinary partial derivative and the spatial covariant derivative, respectively;
$J$ is the determinant of the coframe matrix, see~\eqref{Eq:Jacobian}, and $\CR_{ij}$ is the spatial Ricci tensor with trace $\CR$. (The expression of the Ricci tensor in terms of coframes is left implicit, see \cite{rza1}.)

\subsection{Relativistic Zel'dovich Approximation}\label{SubSec:RZA}

The $3+1$ Lagrangian framework of Einstein's equations consists in considering the (nine functions of the) spatial coframes as the only dynamical variables~\cite{rza1}. The relativistic Lagrangian perturbation theory then only perturbs the coframes, while their first-order member provides the RZA coframes. 
The so linearized Lagrange-Einstein system entitles us to evaluate any other field as a functional of the linear coframe perturbations, leading to nonlinear (functional) expressions for any relevant field.
For instance, in this approximation, the spatial metric is a quadratic form of the deformation field,
\begin{subequations}\label{Eq:RZALineElemGab}
\begin{eqnarray}
g_{ij}&=& G_{ab} \eta^{a}_{~i} \eta^{b}_{~j} 
\\
&=& a^2(t) \left[G_{ij} + G_{ab}  \left(\delta^{a}_{~i}  P^{b}_{~j} + \delta^{b}_{~j} P^{a}_{~i}  + P^{a}_{~i} P^{b}_{~j}  \right) \right] , \qquad
\end{eqnarray}
\end{subequations}
which allows for correctly evaluating distances as well as having the correct light cone structure in generic inhomogeneous matter distributions that correspond to the coframe deformation at a given order.

Since the initial spatial metric is encoded in the Gram's matrix, the deformation field vanishes at some initial time $t_i$,
and we have for the initial data (\textit{cf.} e.g. \cite{rza4}):
\begin{subequations}\label{ICPeq1}
\begin{eqnarray}
\t{P}{a}{i}(t_i) &=& 0 \ ;
\\
\t{\dot{ P}}{a}{i}(t_i)  &\equiv&\t{U}{a}{i} \ ; \quad  {U}_{[ij]}=0 \ ;
\\
\t{\ddot{ P}}{a}{i}(t_i)  &\equiv&\t{W}{a}{i}-2 H (t_i) \t{U}{a}{i} \quad ; \quad {W}_{[ij]}=0 \ .
\end{eqnarray}
\end{subequations}
In the equations above,  $H=\dot a / a$ is the Hubble function,  and the one-form fields $\mathbf{U}^a$ and $\mathbf{W}^a$ are the relativistic generalizations of the initial Newtonian peculiar-velocity and peculiar-acceleration gradients, with coefficients in the exact coordinate basis $\mathbf{d} X^i$ denoted by $U^a_{\ i}$ and $W^a_{\ i}$, respectively. They are subject to the (energy and momentum) constraints of the Einstein equations, Eqs.~\eqref{form_hamiltoncoeff} and~\eqref{form_momcoeff}, here imposed on the initial data:
\begin{subequations}\label{Eq:EMomConstUW}
\begin{eqnarray}
H(t_i) U = -\frac{\CR (t_i)}{4}-W \ ;
\\
(\t{U}{a}{j} \delta^{~i}_{a})_{||i}=(\t{U}{a}{i} \delta^{~i}_{a})_{|j} \ .
\end{eqnarray}
\end{subequations}
Above, $U$ and $W$ denote the traces of the Newtonian peculiar-velocity and -acceleration gradients, respectively, defined below in Eq.~\eqref{Eq:DefTrPWU}. For RZA the general constraint equations are reduced to constraints on initial data due to the space and time-separability of RZA. They therefore hold throughout the evolution even in the approximate regime~\cite{rza1,rza4}. For exact solutions, these constraint equations propagate, also for nonseparable solutions, according to well-known theorems.
 
 
\subsection{Example: The solution for the trace}\label{SubSec:trace}
The solution is separated into spatial and temporal parts.
If we focus on the trace part, the time-dependence is determined from Raychaudhuri's equation~\cite{rza4}, 
\begin{equation}\label{RAY}
 \ddot{P} + 2 H \dot{P} - 4 \pi  \varrho_b(t) P = a^{-3} W \; ,
\end{equation}
where $\varrho_b(t)$ is the background density, and the following abbreviations were and will be used:
\begin{equation}\label{Eq:DefTrPWU}
P\equiv \t{P}{k}{k} = \t{\delta}{k}{a} \t{P}{a}{k} \ ; \; \t{\delta}{k}{a} \t{U}{a}{k} \equiv U\; \ ; \; \t{\delta}{k}{a} \t{W}{a}{k} \equiv W \ .
\end{equation}
Once a background model has been set, the growing and decaying solutions of Eq.~(\ref{RAY}) determine the temporal evolution. 
For an EdS (Einstein-de Sitter) background model, we have:
\begin{eqnarray}
P&=&\frac{3}{5}\Bigg[\left(U t_i +\frac{3}{2}W t_i^2\right) \left(\frac{t}{t_i}\right)^{2/3}
- \nonumber
\\
&{}& \qquad 
\left(U t_i -W t_i^2\right)
\left(\frac{t}{t_i}\right)^{-1}-\frac{5}{2}W t_i^2\Bigg] \, . \;\;\;
\label{Eq:SolPRZA}
\end{eqnarray}
The relativistic correspondence to Zel'dovich's approximation is obtained by subjecting the initial data to the  \textit{slaving condition} $U = W t_i$, \textit{cf.}~\cite{Buchert1989AA,BuchertNonperturbative} for the Newtonian case.

We emphasize that RZA also contains trace-free tensorial parts that include nonperturbative models for gravitational waves. We direct the reader to the detailed analyses in~\cite{rza4}.

\subsection{Functional evaluation}\label{SubSec:FunEvalRZA}

Let us examine in more detail the functional evaluation within RZA. As was pointed out previously, the crucial aspect of the formalism relies on its very architecture, linearizing the deformation field in the coframe set only. All relevant fields are computed from their exact and, in general, nonlinear functional expressions with no further truncations. This extrapolation idea accounts for the intrinsic nonlinearity of the model, encoded in its predicted fields---as in the example of the metric form~\eqref{Eq:RZALineElemGab}, or the implicit functional of the Ricci tensor/scalar in~\eqref{Eq:CoframeRelRicci1}. In this spirit, the nonlinear density field is evaluated through the exact integral of the continuity equation:
\begin{equation}\label{Eq:denFuncGab}
\varrho=\varrho_i J^{-1} \ , \quad \hbox{with} \quad J=\sqrt{g}/\sqrt{G}\ ,
\end{equation}
where the determinant, $J$, is given by
\begin{equation}\label{Eq:Jacobian}
J= \det(\eta^a_{\ i}) = a^3 \big( 1 + {J}^{(1)} + {J}^{(2)}+ {J}^{(3)} \big) \ ,
\end{equation}
and the \textit{peculiar-determinant} is defined through
\begin{equation}
\J  \equiv J /a^3 \ .
\label{Eq:PecJacobian}
\end{equation}
In~\eqref{Eq:Jacobian}, we introduced the principal scalar invariants of the perturbation matrix $\t{P}{a}{i}$,
\bse\label{Eq:Invariants}
\begin{eqnarray}
{J}^{(1)}  & \equiv & \frac{1}{2}\epsilon_{abc}\epsilon^{ijk} P_{\ i}^{a}\delta_{\ j}^{b}\delta_{\ k}^{c} \ ; \label{Eq:invariant1}
\\
{J}^{(2)}  & \equiv & \frac{1}{2}\epsilon_{abc}\epsilon^{ijk} P_{\ i}^{a} P_{\ j}^{b}\delta_{\ k}^{c} \ ; \quad \label{Eq:invariant2}
   \\
{J}^{(3)}  &\!\equiv\!& \frac{1}{6} \t{\epsilon}{-}{abc} \t{\epsilon}{ijk}{} P^{a}_{~i} P^{b}_{~j} P^{c}_{~k}  \ . \label{Eq:invariant3}
\end{eqnarray}
\ese
The expression for the expansion tensor in terms of the coframes follows from the one for the extrinsic curvature, which in a flow-orthogonal foliation of spacetime reads: 
\begin{equation}
\Theta_{ij}=-\mathcal{K}_{ij}=\frac{1}{2} \dot{g}_{ij}
\ .
\end{equation} 
\begin{equation}\label{expansiontensor1}
\t{\Theta}{i}{j}=\t{e}{i}{a} \, \t{\dot{\eta}}{a}{j}\ , \quad \text{with} \quad \t{e}{i}{a} =\frac{1}{2 J}\epsilon_{a b c}\epsilon^{i k l} \eta^{b}_{\ k} \eta^{c}_{\ l} \ .
\end{equation}
Since we are interested in vorticity-free models, the kinematic decomposition of the expansion tensor reduces to
\begin{equation}
\t{\Theta}{i}{j}
=\t{\sigma}{i}{j} \,+\,\frac{1}{3} \Theta \, \t{\delta}{i}{j}
\ ,
\end{equation}
where the expansion scalar, $\Theta=\dot{J}/J$, and the shear tensor, $\sigma_{ij}$, are the trace and trace-free part of the expansion tensor, respectively.

Finally, the three-dimensional spatial curvature and the gravitoelectric and gravitomagnetic parts of the Weyl tensor can be expressed in terms of coframes through the following relations~\cite{rza4}:
\begin{subequations}\label{Eq:ExactCofraRelations}
\begin{eqnarray}
-\CR^i_{\ j} &=& \frac{1}{2 J} \epsilon_{abc} \epsilon^{ikl} \left( \dot{\eta}^a_{\ j} \eta^b_{\ k} \eta^c_{\ l} \right)^{\bdot} - \left( 4 \pi \varrho + \Lambda \right) \delta^i_{\ j} \ ; \label{ricci}
 \qquad
\\
- \frac{\CR}{2} &=& \frac{1}{2J}\epsilon_{abc} \epsilon^{mjk} \dot{\eta}^a_{\ m} \dot{\eta}^b_{\ j} \eta^c_{\ k} - \left( 8\pi  \varrho +  \Lambda  \right) \ ; \label{riccitrace}
\\
&&\frac{1}{2J} \epsilon_{abc}\epsilon^{ik\ell} \ddot{\eta}^a_{\ i} \eta^b_{\ k} \eta^c_{\ \ell}   = \Lambda  - 4 \pi {\varrho} \ ;\label{raych}
\\
-  \t{E}{i}{j} & =&  \frac{1}{2J} \t{\epsilon}{-}{abc} \t{\epsilon}{ikl}{} \addot{a}{j} \a{b}{k} \a{c}{l} + \frac{1}{3} \Big(  4 \pi \varrho -  \Lambda \Big) \t{\delta}{i}{j} \ ;  \label{elec}
\\
- \t{H}{i}{j} &=&  \frac{1}{J} \t{G}{-}{ab} \t{\epsilon}{ikl}{} \big( \adot{a}{j\parallel l} \a{b}{k} + \adot{a}{j} \a{b}{k \parallel l} \big) \ , \label{mag}
\end{eqnarray}
\end{subequations}
where equation (\ref{raych}) follows by taking the trace of equation (\ref{ricci}) and inserting equation (\ref{riccitrace}).\\
See more details about the functional evaluation within RZA in Section~\ref{SubSec:RZAandCPT} and Appendix~\ref{SecApp:FuncEvaluationRZA}.

\section{Szekeres models}\label{Sec:SzekeresModelsintro}

The general line-element of the Szekeres solutions can be cast into the form \cite{Sz75,szafron1977inhomogeneous}:
\begin{equation}\label{Eq:GeneralSzeMetric}
\dd s^2=-\dd t^2+e^{2 \alpha} \dd \z^{2} + e^{2 \beta} \left(\dd \x^{2}+\dd \y^{2}\right),
\end{equation}
where the metric coefficients $\alpha(t,\x,\y,\z)$ and $\beta(t,\x,\y,\z)$  are determined from the Einstein equations, with $\rv=\left(\x,\y,\z\right)$ being the comoving coordinates. For a comprehensive and detailed exposition of the Szekeres models see~\cite{kras1,kras2,BKHC2009}. 

The original Szekeres solution has been reparametrized multiple times. 
However, the whole family can be invariantly defined as an exact solution of the Einstein equations with the following properties~\cite{kras2,wainwright1977characterization}: 
\begin{itemize}
\item[(i)] A geodesic and irrotational dust source.
\item[(ii)] A purely gravitoelectric and Petrov D Weyl tensor.
\item[(iii)] A shear with two equal eigenvalues and degenerate eigensurface coinciding with the one of the Weyl tensor.
\end{itemize}  
This coordinate-independent definition was furnished by Barnes and Rowlingson~\cite{BarnesRowlingson1989SzeInv}.

\subsection{Goode-Wainwright parametrization}\label{GWpara}

In this paper, we use a representation introduced by Goode and Wainwright (GW) in~\cite{GW1,GW2}, which, as we will see below, is well-suited for establishing a formal connection with RZA. 
In Appendix~\ref{SecApp:RelBetweenPars}, we discuss how this parametrization relates to the Szekeres-Szafron's one with a nonvanishing cosmological constant.  
This analysis enhances the GW formulation of class I to include $\Lambda\neq0$, which to the best of our knowledge has thus far not been (formally) considered. The compatibility of class I with the cosmological constant is a natural result, previously used in the literature without formal proof~\cite{IshakPeel2012LargeScale}. Here, we validate and formalize it by analyzing not only the relation between the arbitrary (spatial) functions but also the model's time-evolution (the ultimately $\Lambda$ dynamical contribution). The analogous generalization of class II is due to Meures and Bruni~\cite{SzeLamddanot0Meures}.

In the GW representation, the Szekeres line-element reads:
\begin{equation}\label{Eq:SzeMetricGW}
\dd s^2=-\dd t^2 + \s^2  \left(\G ^2 \W^2 \dd \z^{2} + e^{2 \nu}  \left( \dd \x^{2} + \dd \y^{2}\right) \right) \ ,
\end{equation}
where the metric function $S(t,\z)$ satisfies
\begin{equation}\label{Eq:FriedmannLikeEqn}
\dot{\s}^2=-k_0 + \frac{2 \mu}{\s} +\frac{\Lambda}{3}\s^2\ .
\end{equation}
Here, $k_0=0,\pm1$, while $\mu=\mu(\z)$ is arbitrary. In general, $\mu>0$ is needed to have a well-defined FLRW limit.
Next, $\G(\z,\x,\y)$ is given by
\begin{eqnarray}
\label{Eq:GDef}
\G&=&\A(\rv)-\F(t,\z)\nonumber
\\
&=&\A(\rv)-\beta_{+} f_{+}-\beta_{-} f_{-} \ .
\end{eqnarray}
$\mathcal{A}$, $e^{2\nu}$, $\mathcal{W}$, $\beta_{+}(\z)$ and $\beta_{-}(\z)$ differ for each class, and $f_{+}$ and $f_{-}$ are the growing and decaying solutions of
\begin{equation}\label{Eq:Ftt}
\ddot{\F}+ 2\frac{\dot{\s}}{\s}\dot{\F}-\frac{3\mu}{\s^3}\F=0 \ ,
\end{equation}
which can be traced back to the Raychaudhuri equation.

The energy-density takes the following simple form:
\begin{equation}\label{Eq:Rho2ClassesGenEq}
8 \pi \varrho(t,\rv)=\frac{6 \mu \A}{\s^3 \G}=\frac{6 \mu}{\s^3}\left(1+\frac{\F}{\G}\right) \ .
\end{equation}

The solutions of~\eqref{Eq:FriedmannLikeEqn} and~\eqref{Eq:Ftt} can be expressed in parametric form in the cases of a vanishing cosmological constant (see Appendix~\ref{App:ParametricSols}), and for class II with $k_0=0$, but $\Lambda\neq0$, see~\cite{SzeLamddanot0Meures}.

The model is separated into two classes as follows:
\subsubsection*{Class I , $\beta_{,\z}\neq0$ in~\eqref{Eq:GeneralSzeMetric}}
\noindent
For this class,
\begin{subequations}
\begin{eqnarray}
\s=\s(t,\z)\ , \quad  \hbox{with} \quad \s_{,\z}\ne0 \ ,
\\
f_{\pm}=f_{\pm}(t,\z)\ , \quad \mathcal{T}=\mathcal{T}(\z)\ , \quad \mu=\mu(\z) \ ,
\end{eqnarray}
and
\begin{eqnarray}
e^\nu&=&f(\z)\Big[c_0(\z)(\x^2+\y^2) \nonumber
\\
&{}& \quad+2c_1(\z)\x+2c_2(\z)\y+c_3(\z)\Big]^{-1} ; \label{Eq:eToNuGW}
\end{eqnarray} 
while $f(\z)$ is completely arbitrary, the $c_i$ functions are subject to the conditions:\footnote{
As defined in~\eqref{ClassIBeta}, our function $\beta_-$ differs by a factor of $\mu$ from its equivalent in the original GW parametrization, $\beta_-=f  \mathcal{T}_{,\z}/(6\mu)$.
Since the solutions of Eq.~\eqref{Eq:Ftt} are determined up to a multiplicative function of $\z$, we have the freedom to choose that function under the criterion of simplicity. 
The physics of  the solution  is not  contained in the function $\beta_-$ alone, but in the  expression for  scale factor $\G=\A-f_-\beta_--\beta_+ f_+$. See Ref. 20 in~\cite{GW1} and Appendix~\ref{SecApp:RelBetweenPars}.
}
\begin{eqnarray}
&{}& c_0 c_3-c_1^2-c_2^2=\epsilon/4\ ,\,\quad \quad \epsilon=0\ ,\pm 1 \ ; \label{SubEq:Rel-ci}
\\
&{}& \A=f\nu_{,\z}-k_0\beta_+ \ , \quad \;  \W^2=(\epsilon-k_0 f^2)^{-1} \ ; \; \label{Eq:AandWclassIGW}
\\ 
&{}& \beta_+=-k_0 f \mu_{,\z}/(3\mu)\ ,\, \quad \beta_-=f  \mathcal{T}_{,\z} \ .
\label{ClassIBeta}
\end{eqnarray}
 \end{subequations}

The present parametrization was originally formulated by Goode and Wainwright by assuming $\Lambda=0$. 
Since then, the GW parametrization of class I has been restricted to the case without a cosmological constant.
In Appendix~\ref{SecApp:RelBetweenPars}, we provide the formal proof that the GW parametrization is valid for $\Lambda\neq0$ as well, filling a gap in the literature on this topic.

Szekeres models predict an inhomogeneous initial (past) singularity, the ``big bang time'', $\mathcal{T}(\z)$, one of the free functions of the model. Note that it is not correct to associate this singularity with the physical Big Bang since these dust cosmological models are not valid in a radiation-dominated era. 
Due to the relation between $\mathcal{T}_{,\z}$ and the decaying mode of structure~\cite{CharGrowingDecaying}, some authors assume a simultaneous bang time condition to have a ``purely growing mode"~\cite{Nwankwo_2011,IshakPRL2013,IshakNwankwo2008}.  The absence of the decaying mode is motivated by its negligible contribution in the matter-dominated era.\footnote{For RZA this is realized by the alignment between $\mathbf{U}^a$ and $\mathbf{W}^a$ after recombination (see the \textit{slaving condition} imposed on initial data that leads to the strict absence of the decaying mode in the matter-dominated regime, section~\ref{SubSec:trace}). 
The analysis of the radiation-dominated epoch shows that matter perturbations grow logarithmically during this phase leading to this alignment or \textit{slaving} of the motion to the gravitational field~\cite{Padmanabhan}.} However, for the sake of mathematical generality, we will not make any restricting assumption on $\mathcal{T}$ in the present paper. 
 
\subsubsection*{Class II , $\beta_{,\z}=0$ in~\eqref{Eq:GeneralSzeMetric}}
\noindent
This class has a much simpler mathematical structure than the previous one:
\begin{subequations}\label{Eq:SolMetricClassII}
\begin{eqnarray}
\s=\s(t)\ , \quad f_{\pm}=f_{\pm}(t)\ , \quad \mathcal{T}, \mu=\hbox{const.} \ ;
\\
 k_0=0,\pm1\ ,\quad \W=1 \ ;
 \\
e^\nu=\left[1+\frac{k_0}{4}(\x^2+\y^2)\right]^{-1} \ ,
\label{Eq:etonuClassII}
\end{eqnarray}
with
\begin{equation}\label{Eq:DefAclassII}
  \A = \left\{ 
  \begin{array}{l l}
    e^\nu \Big[ c_0(\z) \left(1-\frac{k_0}{4}(\x^2+\y^2)\right) + c_1(\z)\x \\
    \hspace{.9cm}  +c_2(\z)\y \Big]-k_0\beta_{+}\ ,\; \hbox{for} \; k_0=\pm 1\ ;
    \\
    c_0(\z)+c_1(\z)\x+c_2(\z)\y
    \\
    \hspace{.9cm}-\beta_{+} (\z) (\x^2+\y^2)/2 \ , \;  \hbox{for} \; k_0=0 \ ;
  \end{array} \right.
\end{equation}
\end{subequations}
$c_i$ and $\beta_{\pm}$ are arbitrary functions of $\z$. 
Here, the constant $\mu$ satisfies
\begin{equation}
 3 \mu=4 \pi \varrho_b(t_i) \ ,
 \end{equation} 
and $\T$ can be set to zero without loss of generality,\footnote{This assumption does not impose any restriction on the decreasing mode.  For class II, $\beta_-$ is an arbitrary function, in contrast to class I where $\beta_- \propto \mathcal{T}_{,\z}$.} setting the initial singularity at $t=0$
(for cosmological applications we are usually interested in $t\geq t_i>0$).
Then, Eq.~\eqref{Eq:Rho2ClassesGenEq} can be rewritten as follows:
\begin{equation}\label{Eq:RhoClassII}
\varrho(t,\rv)= \varrho_b(t)\left(1+\frac{\F}{\G}\right) \ ,
\end{equation}
where we can identify the background density ($\varrho_b$) and its exact ``perturbation'' (deviation: $\delta=\varrho/\varrho_b-1=\F/\G$, the usual variable of cosmological perturbation theory).

\hfill

As was noted by Goode and Wainwright~\cite{GW1}, one of the remarkable properties of the Szekeres solutions in this representation is the role of equations \eqref{Eq:FriedmannLikeEqn} and~\eqref{Eq:Ftt} governing the evolution. The first one is the well-known Friedmann equation (for fixed $\z$ in class I), while the second one is the same equation that in standard linear perturbation theory leads to the growing and decaying modes. The latter admits a first integral in the form~\cite{GW1,SzeLamddanot0Meures}:
\begin{equation}\label{Eq:1st-IntFtt}
\frac{\dot \s}{\s} \dot{\F}-\frac{\left(k_0 -3 \mu/\s \right)}{\s^{2}}\F=\frac{\mathcal{C}}{\s^2} \ , \quad \text{with} \quad \mathcal{C}=\beta_+ \ ,
\end{equation}
where the presence of an inhomogeneous term in the differential equation is due to the growing contribution of $\F$, $\beta_+ f_+$. 
The constant $\mathcal{C}$ arises as residual freedom of the integration of~\eqref{Eq:Ftt} and its value cannot be determined from the line-element alone. The condition $\mathcal{C}=\beta_+$ is more than a mathematical simplification, it can be obtained from the Einstein equations.\footnote{As was shown in~\cite{SzeLamddanot0Meures}, $\mathcal{C}=\beta_+$ can be obtained directly from the Einstein equations, and the term $\mathcal{C}/\s^2$ represents an inhomogeneous curvature. Although most of the analysis was restricted to the case $k_0=0$ of the class II models, the steps leading to~\eqref{Eq:1st-IntFtt} in Appendix B of~\cite{SzeLamddanot0Meures} can be easily generalized to the whole family of Szekeres models. See also~\cite{rza3} for similar first integrals of Eq.~\eqref{RAY} in RZA and~\cite{BruniEtal2013} for a perturbative analysis.}

Strictly speaking, the Goode and Wainwright formulation does not fully cover the Szekeres models~\cite{GW1}. Since the success of their reparametrization relies on expressing the line-element in terms of the growing and decaying linear modes on an FLRW universe model, solutions with an associated vacuum background ($\mu=0$) are left out of the description. Unless otherwise specified, in this paper, we will refer to Szekeres class II as those solutions included in the GW formulation, excluding the ``PII'' and ``HIII'' cases of~\cite{BonnorTomimura1976evolution}, which, however, exhibit interesting mathematical properties~\cite{SenovillaVera2000PII}. 

\subsection{FLRW limit and Cartesian coordinates}

We first note that $\rv=\left(\x,\y,\z\right)$ is the coordinate system in which the Szekeres spatial metric is diagonal, but they are not coordinates of any of the standard representations of the FLRW solution. 
In fact, the FLRW models emerge in an unfamiliar form in this coordinates when $\beta_{+}=\beta_{-}=0$ (necessary and sufficient conditions for the Friedmann limit)~\cite{GW1,kras1,kras2}. In $\rv$-coordinates, the FLRW line-element is described by the following coframe set:
\begin{equation}\label{Eq:RWCofXiClassI}
\overline{\boldsymbol{\eta}}^{1} =a\, e^\nu {\bf d} \x
\ ;
\;
\overline{\boldsymbol{\eta}}^{2}=a\, e^\nu {\bf d} \y
\ ;
\;
\overline{\boldsymbol{\eta}}^{3}=a \, f \, \W \,\nu_{,\z} {\bf d} \z \ ,
\end{equation}
for class I; and
\begin{equation}\label{Eq:RWCofXi}
\overline{\boldsymbol{\eta}}^{1} = a\, e^\nu\, {\bf d} \x
\ ;
\quad
\overline{\boldsymbol{\eta}}^{2}= a \,e^\nu\, {\bf d} \y
\ ;
\quad
\overline{\boldsymbol{\eta}}^{3}=a \,\bar{\A} \, {\bf d}  \z \ ,
\end{equation}
for class II.
In the above expressions, 
\begin{equation}
 \bar{\A}\equiv\A{\big|}_{\beta_{\pm}=0} \ ,
\end{equation}
and we have replaced $\s$ by the FLRW scale factor, $a(t)$. 

An appropriate selection of the arbitrary functions can lead to the FLRW limit in a more specific coordinate system.
In particular, the choices~\cite{kras1,kras2}
\begin{equation}\label{Eq:RWclassI}
c_1=c_2=0 \ ; \quad c_3=4 \, c_0=1 \ ; \quad f=\z=r \ , 
\end{equation}
for class I, and,
\begin{equation}\label{Eq:RWclassII}
c_0=c_2=0 \ ;  \qquad c_1=1 \ ,
\end{equation}
for class II, reduce the line-element to forms which can be transformed to Cartesian coordinates by the changes of coordinates~\eqref{App:TransToCartII} and~\eqref{App:TransToCartI} (see Appendix~\ref{App:FLRW-Cartesians}). Thus, the FLRW coframe set takes the trivial form $\overline{\boldsymbol{\eta}}^{i} = a(t) \, {\bf d} X^i \; .$

One may wonder why not working in Cartesian coordinates from the very beginning. There are two main reasons for not doing so: first, in the diagonal coordinate system, we can undertake the analysis without additional nondiagonal terms hindering the interpretation of the results. Second, we do not have a global transformation to take the FLRW model in the GW representation to Cartesian coordinates. Instead, we need to impose additional assumptions (equations~\eqref{Eq:RWclassI} and~\eqref{Eq:RWclassII}). Although we argue that these assumptions can be seen as a particular choice of coordinates of the FLRW limit, essential physics could be lost. For instance, the Szekeres dipole of class I is related to the functions $c_i$; setting them to fixed values imposes significant restrictions on the dipolar anisotropy. 

\subsection{One motivation of this work: Kasai's statement}\label{KasaiDiscussion}

So far we have presented the most fundamental aspects of RZA and Szekeres solutions by separating the respective classes. Masumi Kasai~\cite{Kasai1995} pointed out a relation, suggesting that the trace-part of RZA is contained as a particular case within the class II of Szekeres models with $k_0=0$. Particularly, Kasai states that this subfamily satisfies ``the linearized constraint equation'' of RZA, leaving open the questions of whether such a relation holds exactly or whether it is strictly linear, and also whether there are other subcases of the Szekeres solutions contained within RZA. 
Kasai has also restricted the RZA model to the consideration of the coframe deformation, {\it i.e.} he does not functionally extrapolate variables other than the density, {\it e.g.} he proposes to linearize the metric functional~\eqref{Eq:RZALineElemGab}.
In the following sections, we will address these issues by a direct comparison of the line-elements.

\section{Szekeres exact solutions as Relativistic Lagrangian perturbations}
\label{Recipe}

To relate the Relativistic Lagrangian formalism to the Szekeres exact solutions, 
we will take the following steps:
%
%
\begin{enumerate}
\item %
Find the set of coframes in the orthonormal Cartan basis $\widetilde{\eta}^{ a}_{~i}$,
where the metric is diagonal,
\begin{equation}\label{Eq:tildeCoframes}
^{(3)}\mathbf{g}=\delta_{ab} \widetilde{\boldsymbol{\eta}}^{ a}\otimes\widetilde{\boldsymbol{\eta}}^{ b} \ .
\end{equation}
Here, these coframes can be considered, 
without loss of generality, diagonal as well. 
\item  %
Obtain the Gram's matrix, identified with the initial metric, \textit{cf.} Appendix \ref{SecApp:FuncEvaluationRZA}:
\begin{equation}\label{Eq:GramsMatrix}
G_{ab}=\delta_{cd} \, \widetilde{\eta}^{ c}_{\ a}\big|_{t=t_i}\widetilde{\eta}^{ d}_{\ b}\big|_{t=t_i} \ .
\end{equation}
Then, the coframes can be formally rewritten as in~\eqref{Eq:CoframesGab}, $\eta^{a}_{~i}=a(t) \left(\delta^{a}_{~i}+P^a_{~i}\right)$,
\begin{equation}\label{Eq:LineElement_f2}
^{(3)}\mathbf{g}=\delta_{cd} \, \widetilde{\boldsymbol{\eta}}^{c}\otimes\widetilde{\boldsymbol{\eta}}^{d} 
= G_{ab} \boldsymbol{\eta}^{ a}\otimes \boldsymbol{\eta}^{ b} 
\ .
\end{equation}
%
\item  %
Split the spatial Szekeres metric into the initial metric ($G_{ij}$) and its exact deviation 
(${h}_{ij}$):
\begin{equation}
g_{ij}\equiv a^2 (t) \gamma_{ij} \equiv a^2 (t) \left(G_{ij} 
+ {h}_{ij} \right).
\label{RZAansatzSzeII}
\end{equation}
Here, ${h}_{ij}$ vanishes at the initial time $t_i$.
\item  %
Solve the equations for the deformation field,
\begin{equation}\label{GammaijSze} 
{h}_{ij}\equiv G_{ab} \left(\delta^{a}_{\ i}  P^{b}_{\ j} + \delta^{b}_{~j} P^{a}_{\ i}  + P^{a}_{~i} P^{b}_{\ j} \right) \ ,
\end{equation}
resulting from equating the Szekeres and RZA metric components in the $\rv-$coordinates.
%
%
\end{enumerate}
%
Note that Step $1$ leads to the coframes in the coordinates $\rv$, where the FLRW line-element takes an unfamiliar form. However, although the relationship between these and the Cartesian coordinates is not straightforward, there exists a spatial transformation taking the line-element from one coordinate system to another, with the 
fluid's $4$-velocity being in both cases a vector normal to the spatial hypersurfaces. 

To obtain the deformation field of the Szekeres model, let us follow step by step the recipe proposed above. Equation~\eqref{Eq:tildeCoframes} provides a link between the coframes and the Szekeres line-element, 
\begin{equation}\label{Eq:SzeCofXi}
\left[ g_{ii} \right]=[ \left(\tilde{\eta}^{ i}_{~i}\right)^2 ]  \quad \hbox{in $\rv$-coordinates}\ ,
\end{equation}
where ``$[\;]$'' denotes the component of the element with no summation implied, and the remaining (off-diagonal) 
equations are identically satisfied with $\eta^j_{~i}=0$ ($j\neq i$). Then, 
\begin{equation}\label{Eq:SecCofXi}
\tilde{\boldsymbol{\eta}}^{1} = \s e^\nu {\bf d}\x
\ ;
\quad
\tilde{\boldsymbol{\eta}}^{2}=\s  e^\nu {\bf d} \y
\ ;
\quad
\tilde{\boldsymbol{\eta}}^{3}=\s \G \W {\bf d} \z \ ,
\end{equation}
where $S$, $e^{\nu}$, $\G$ and $\W$ were given in Section~\ref{Sec:SzekeresModelsintro} for each class.

This is the farthest we can go without splitting the analysis into the classes. 
The connection with RZA requires a time-dependent conformal FLRW scale-factor, Eq.~\eqref{Eq:RZALineElemGab}. While in class II such a scale factor emerges naturally, in class I, the conformal metric function depends on both temporal and spatial coordinates, which notably hinders the connection between both solutions, but also hints to a possible generalization of RZA.

\section{Szekeres Class II and RZA}\label{Sec:ClassIIRZA}

We begin by examining the models of class II, because their simple mathematical structure allows for a straightforward connection with RZA. 
In these spacetimes, the surfaces $\left\{(t, \z)=\text{const.}\right\}$ have constant curvature; however, the plane, spherical or hyperbolic symmetry is lost for an unrestricted set of the arbitrary functions, leading to the characteristic absence of killing vectors in the Szekeres solutions, but pointing to certain quasisymmetries~\cite{Bonnor1977,kras1}.
In what follows, we will denote $\s \equiv a(t)$ to make more explicit its relation to the Friedmannian scale factor. Also, the time of the initial singularity (``bang time'') will be set to zero, $\T=0$.\\

For class II, the coframes~\eqref{Eq:SecCofXi} take the following form:
\begin{subequations}
\begin{eqnarray}
\tilde{\boldsymbol{\eta}}^{1} &=& a(t) e^\nu {\bf d}\x
\ ;
\\
\tilde{\boldsymbol{\eta}}^{2}&=&a(t)  e^\nu {\bf d} \y
\ ;
\\
\tilde{\boldsymbol{\eta}}^{3}&=&a(t) \G {\bf d} \z = a(t)  \left(\A-\F\right) {\bf d} \z 
\nonumber\\ 
&=& a(t)  \left(\widetilde{\A}-\widetilde{\F}\right)  {\bf d} \z\ , 
\label{coframesII}
\end{eqnarray}
\end{subequations}
where 
\bse
\begin{eqnarray}
\widetilde{\A}\equiv \A-\F_i  \ , \qquad \widetilde{\F}\equiv \F-\F_i \ , 
\\
\text{with} \qquad \F_i=\F(t_i, \z) \ .
\end{eqnarray}
\ese
Next, Gram's matrix is determined by substituting the previous expressions into~\eqref{Eq:GramsMatrix}, 
\begin{equation}\label{Eq:GabClassIISol}
G_{ab}=\hbox{Diag}\left[ e^{2\nu},e^{2\nu}, \widetilde{\A}^2\right] \ ,
\end{equation}
with $a(t_i)=1$.

By comparing the line-elements~\eqref{Eq:RZALineElemGab} and~\eqref{Eq:SzeMetricGW} and using~\eqref{GammaijSze}, we find that the only nontrivial component of ${h}_{ij}$ is given by
\begin{eqnarray}
{h}_{33} =-2 \widetilde{\A} \widetilde{\F} +\widetilde{\F}^2
&=& {G}_{ab} \left(\delta^{a}_{~3}  P^{b}_{~3} + \delta^{b}_{~3} P^{a}_{~3}  + P^{a}_{~3} P^{b}_{~3}\right)  \nonumber
\\
 &=& 2 \widetilde{\A}^2 P^{3}_{\ 3} +  \widetilde{\A}^2 \left( P^{3}_{\ 3} \right)^2 \ , 
\end{eqnarray}
which yields:
\begin{equation}\label{Eq:P33eqn1}
P^{3}_{\ 3} = -\widetilde{\F}/\widetilde{\A} \ .
\end{equation}
Hence, $P^{3}_{\ 3}$ is the only nonvanishing element of the deformation field in these coordinates.
So far, this is just a convenient ansatz. The connection with RZA not only involves the decomposition of the spatial metric into a bilinear quadratic form as in \eqref{Eq:RZALineElemGab}, but also the matching of the temporal evolutions. To address this issue, we first note that
\begin{equation}\label{Eq:PsEvolClassII}
P=\t{P}{3}{3}=\gamma_+ f_+ - \gamma_-  f_-   -  \left(  \gamma_+ f_{+}(t_i) -  \gamma_- f_{-}(t_i) \right) \ ,
\end{equation}
where we have defined $\gamma_\pm\equiv\mp\beta_\pm/\widetilde{\A}$. 

In this way, the temporal evolution of the Szekeres deformation field is entirely contained in the functions $f_{\pm}(t)$, the growing and decaying solutions of~\eqref{Eq:Ftt}. It is remarkable that this equation differs from Eq.~\eqref{RAY} only by a nonhomogeneous term. However, from the theory of second-order ODEs,  the solution of the nonhomogeneous equation~\eqref{RAY} can be expressed as the sum of the general solution of the homogeneous equation and a particular solution of the nonhomogeneous one, $f_p$,
\begin{equation}\label{Eq:GenPRZA}
P=\zeta_+ f_+ + \zeta_- f_-  + f_p \ .
\end{equation}
The functions $\zeta_\pm$ are constant in time, but can depend on the spatial coordinates.
The choice for $f_p$ that matches RZA initial data is:
\begin{equation}\label{Eq:GenSolf_par}
f_p=-W/\left(4 \pi \varrho_b(t_i)\right) \ ,
\end{equation}
where $W$ is the trace of the generalized Newtonian peculiar-acceleration gradient, defined further below in terms of Szekeres functions.

By a direct comparison of the trace-part evolution of the deformation fields, Eq.~\eqref{ICPeq1} for RZA and~\eqref{Eq:GenPRZA}-\eqref{Eq:GenSolf_par} for Szekeres, we find:
\begin{subequations}
\begin{eqnarray}
\zeta_+ =  \gamma_+=-\beta_+/\widetilde{\A} \ , \quad \zeta_-= -\gamma_- =-\beta_-/\widetilde{\A} \ ;
\\
 \gamma_+ f_{+}(t_i) -  \gamma_- f_{-}(t_i) =W/\left(4 \pi  \varrho_b(t_i)\right) \ ; \quad
 \\
  \gamma_+ \dot{f}_+(t_i) -  \gamma_- \dot{f}_-(t_i)  =U \ ; \quad
  \\
   \gamma_+ \ddot{f}_+(t_i) -  \gamma_-  \ddot{f}_-(t_i)  =W -2 H(t_i) U \ . \quad
\end{eqnarray}
\end{subequations}
The only nontrivial components of the generalized initial peculiar-velocity and peculiar-acceleration gradients are $U^{3}_{\ 3}$ and $W^{3}_{\ 3}$, so that $U=U^{3}_{\ 3}$ and $W=W^{3}_{\ 3}$.
These initial data strictly obey the constraint equations~\eqref{Eq:EMomConstUW}, which are propagated during the temporal evolution since the Szekeres models are exact solutions.

At last, we are in the position to discuss the reasoning behind the approach followed in this paper. The GW formulation of the Szekeres solutions resembles the usual formalism of cosmological perturbations: the solution splits into a background and (exact) deviations thereof. Such deviations obey the well-known equation for the evolution of the linear modes on an FLRW background. This feature was and will also be here exploited to establish a formal relation between RZA and the Szekeres solutions. 

The current analysis can be regarded as a reinterpretation of the Szekeres solutions in terms of the initial metric perturbations (Gram's matrix) and a deformation field: the RZA model variables. Aside from the associated background evolution, the nontrivial dynamics of Szekeres class II is contained in the metric function $\G=\A-\F$, where $\F\equiv0$ unequivocally determines the Friedmann limit. This intuitively justifies why we have only one nontrivial component of the deformation field. The results obtained in this section show that RZA contains the whole class II of Szekeres models as a particular case, and also provide a reinterpretation of the Szekeres arbitrary functions in terms of the generalized initial Newtonian peculiar-velocity and -acceleration gradients.

It is worth noting that no assumption has been made on the associated background, making the analysis valid for any background, either curved, EdS, or $\Lambda$CDM. Our discussion proves and generalizes the remark made by Kasai in~\cite{Kasai1995}, where he states that the Szekeres solution of class II with an associated EdS background satisfies ``the linearized constraint equation'' of RZA. 


\subsubsection*{Example: Einstein-de Sitter background}

For an Einstein-de Sitter associated background, the conformal scale factor and the growing and decaying solutions of~\eqref{Eq:Ftt} are given by
\begin{equation}\label{Eq:EdS_fp-fm}
a(t)=\left(t\big/t_i\right)^{\frac{2}{3}} \ ,\quad f_+=\left(t / t_i\right)^\frac{2}{3} \ , \quad f_-=\left(t / t_i\right)^{-1}\ .
\end{equation}
From the Szekeres deformation field 
we can identify the arbitrary functions with the initial 
traces of the generalized Newtonian peculiar-velocity and peculiar-acceleration gradients,
\bse
\begin{eqnarray}
\frac{2 \gamma_+ + 3\gamma_-}{3 t_i}=U=U^3_{\ 3} \ ;
\\
 \frac{2 \left( \gamma_+ - \gamma_- \right)}{3 t_i^2}=W=W^3_{\ 3} \ .
\end{eqnarray}
\ese
In Eq.~\eqref{Eq:EdS_fp-fm}, the constant $t_i$ remains undetermined as a residual freedom of the integration of Eq.~\eqref{Eq:Ftt}.
In particular, the first integral~\eqref{Eq:1st-IntFtt} sets the arbitrary constant $t_i$ to $t_i^{\,2}=10/9$.

\subsection{A note on admissible initial data for class II}\label{SubSecNoteAdm}

In perturbative settings, where deviations off a preassumed background are imposed, such deviations have to obey certain integral constraints. In order to make sense to speak of a background, the deviations thereof have to average out on some large scale of homogeneity $L_H$. 

We may begin to look at the following constraint on the RZA coframe set \eqref{Eq:CoframesGab}: 
\begin{equation}
\int_{\cD_H} \t{P}{a}{i} \, J \dd^3 {\mathbf X} \; = \; 0 \ \ ;\ \  J = \frac{\sqrt{g}}{\sqrt{G}} \ ; \ g({\mathbf X,t_i})\equiv G \ ,
\label{Eq:EclideanIntConst}
\end{equation}
where $\mathcal{D}_H$ denotes a spatial domain of averaging corresponding to the homogeneity scale $L_H$, $\dd^3 {\mathbf X}=\sqrt{G} \,\dd^3 \rv$, and $J \,\dd^3 {\mathbf X}$  is the Riemannian volume element. 

Applied to the class II of Szekeres models, it turns out 
that the integral constraint~\eqref{Eq:EclideanIntConst}, which is motivated by 
perturbations on a flat space, $J=1$ (see \cite[Eq. (26c)]{ehlersbuchert}), is not correct in 
a Riemannian space. One can convince oneself of this
fact by looking at the trace of the integrand and noticing that it splits into $a^3 \int_{\cD_H} P \dd^3 \mathbf{X}$ and $a^3  \int_{\cD_H} P^2 \dd^3 \mathbf{X}$. The final expression is a sum of linearly independent functions that vanishes if only if the deformation field vanishes identically (valid in any coordinate system). 

The search for integral constraints on the initial data takes us to the backreaction problem, {\it i.e.} the property that deviations average out on the scale $L_H$ is equivalent to proving that the kinematical backreaction $\CQ_{\cD_H}$ (see Appendix~\ref{App:BackAveEqns}) vanishes, leading to our first lemma. 
\begin{lemma}\label{Lem:lemma1}
Given a compact domain of homogeneity, $\cD_H$, the vanishing of $\averageH{\dot P/\mathfrak{J}}$ is a necessary and sufficient condition for the absence of backreaction on $\cD_H$:
\begin{equation}
\CQ_{\cD_H}=0 \quad \Leftrightarrow \quad \averageH{\frac{\dot P}{\mathfrak{J}}}= \ 0 \quad \text{(for class II)} \ .
\end{equation}
\end{lemma}

Here, we have introduced the spatial average of a scalar-valued field on a compact domain $\cD$, which is defined as~\cite{Buchert2000GERG}:
\begin{subequations}
\begin{eqnarray}
\average{\Psi}&=&\frac{1}{V_{\cD}} \int_{\cD} \Psi \, J \, \dd^3 {\mathbf X} \ ,
\\
&\hbox{with}& \quad V_{\cD}= \int_{\cD}  J \, \dd^3 {\mathbf X} \ .
\end{eqnarray}
\end{subequations}
\noindent  
{\bf Proof.}  The backreaction source term (on an arbitrary compact domain $\cD$) is given by~\cite{Buchert2000GERG}
\begin{subequations}
\begin{eqnarray}
\CQ_{\cD}&\equiv& 2 \average{\inII}-\frac{2}{3} \average{\inI}^2  \label{Eq:QdInv-a}
\\
&=& \frac{2}{3} \left( \average{\Theta^2} - \average{\Theta}^2\right)-2 \average{\sigma^2} \ , \label{Eq:QdScal-b}
\end{eqnarray}
\end{subequations}
where $\sigma^2 \equiv \frac{1}{2}\t{\sigma}{i}{j}\t{\sigma}{j}{i}$ is the rate of shear, and 
$\Theta$ and $\t{\sigma}{i}{j}$ are the expansion scalar and shear tensor. 
Substituting their functional expressions (Eq.~\eqref{Eq:ThetaClassII} and~\eqref{Eq:ShearClassII} further below) we obtain,  \textit{cf.} Appendix~\ref{App:ProofTheor2}:
\begin{equation}
\label{offsetbackreaction}
\CQ_{\cD}=-6 \average{\Sigma}^2  \leqslant 0,  \quad \text{(for class II)} \ ,
\end{equation}
with $\Sigma \equiv -\left(1/3\right) \, (\dot P/\mathfrak{J})$. {\tiny $\blacksquare$}

This lemma provides us with a workable integral constraint for a domain of homogeneity, $\cD \equiv \cD_H$: 
\begin{lemma}\label{Lem:lemma2}
The vanishing of the average of $\dot P/\mathfrak{J}$ on the homogeneity scale $L_H$ is a necessary and sufficient condition for the vanishing of $ P/\mathfrak{J}$ on $\cD_H$:
\begin{equation}\label{Eq:ConsIfonlyIfClassII}
  \averageH{\frac{ P}{\mathfrak{J}}}=0 \quad \Leftrightarrow \quad \averageH{\frac{\dot P}{\mathfrak{J}}}=0  
  \quad 
 \text{(for class II)}\ .
\end{equation}
\end{lemma}
\noindent  
{\bf Proof.} For this class, all relevant functions split into a temporal and spatial dependence; then, the integral constraints $\averageH{P/\J}=0$ and $\langle{\dot P/\J}\rangle_{\cD_H}=0$ can be written as $I_+ q_+(t)+ I_- q_-(t)=0$ and $I_+ \dot{q}_+(t) + I_- \dot{q}_-(t)=0$, respectively, where  $I_\pm$ are constants arising from the spatial integration. Since $q_\pm$ and $\dot{q}_\pm$ are linearly independent functions, 
each constraint requires that $I_\pm=0$, which ensures their simultaneous fulfillment and proves the lemma. {\tiny $\blacksquare$}

For an alternative and more formal proof based on the commutation rule, see Appendix~\ref{App:ProofTheor2}. 

These integral constraints can be traced back to the assumption of setting the average values of density and curvature equal to their FLRW values. We will show that their subsequent propagation preserves these properties. We formalize this in the Lemma below, providing a physical meaning to the vanishing of kinematical backreaction for class II. 
%
%
\begin{lemma}\label{Lem:ConsRho-3R}
The integral constraints~\eqref{Eq:ConsIfonlyIfClassII} are equivalent to the following conditions:
\bse\label{Eqns:ConClassIILem}
\begin{eqnarray}
&{}& \text{{\it(i)}} \quad \averageH{\varrho}=\varrho_b(t) \ ; \label{Eq:MassConClassIILem}
\\
&{}& \text{{\it(ii)}} \quad \averageH{\CR}=6 \frac{ k_0}{a^2} \ . \label{Eq:CurvConClassIILem}
\end{eqnarray}
\ese
\end{lemma}
%
%
This lemma is proved in Appendix~\ref{AppSec:ProofOfLemmata3-4}.  In the above equations, $\varrho_b(t)$ and $k_0$ are the Friedmannian quantities appearing in the equation for the conformal scale factor.

The conservation law for  the curvature is a natural result of the vanishing of kinematical backreaction. 
Setting the initial averaged curvature (on $\cD_H$) equal to the initial FLRW constant value,  its conservation law follows from $\CQ_{\cD_H}=0$ and the integrability condition, Eq.~\eqref{AppEq:Integrability}. 

There is the freedom to assume corresponding offsets in the model initial data, but if these offsets are set to zero, the average values of density and curvature equal the FLRW values at all times as a result of the conservation laws for the average density and average curvature. However, if we adopt the point of view that, at early stages, the Universe can be considered as a perturbation of an FLRW model, such offsets will not be justified. 

\begin{proposition}\label{Prop:TrueIntConst}
The integral constraint~\eqref{Eq:EclideanIntConst} finds its generalization to the Szekeres class II geometry in
\begin{equation}\label{Eq:TrueIntConst}
\Bigl\langle{\frac{P}{\J}}\Bigr\rangle_{\cD_H}\,=\,0    \quad 
 \text{(for class II)} \  .
\end{equation}
\end{proposition}
If the deviation functions are integrable, {\it i.e.} if they are exact one-form fields, ${\mathbf P}^a = \mathbf{d} F^a$,
then this integration results in boundary terms. Such a procedure is commonly imposed in Newtonian simulations, where the deviation fields $F^a$ are subject to periodic boundary conditions, {\it i.e.} assuming a spatial 3-torus topology on the scale $L_H$ \cite{BuchertEhlers}. Integration is then over the whole boundary-free space $\mathcal{D}_H =:\Sigma$.
Nonvanishing averages in the interior of the 3-torus model are commonly called \textit{cosmic variance}. 

The integrability of our deformation field can be easily verified, if we take into account that the general expression, 
\begin{equation}
\bm{P}^a=P^a_{~i} \, \mathbf{d}X^i \ ,
\end{equation}
can be reduced to
\begin{eqnarray}
\bm{P}^3&=&\t{P}{3}{3} \, \sqrt{G_{3 3}}\,\mathbf{d}\z= -\left(\F-\F_i\right) \mathbf{d} \z 
\nonumber \\
&=&-\bigg\{\beta_+(\z)\left(f_+(t)-f_+(t_i)\right)
\nonumber
\\
&{}&+\beta_-(\z)\left(f_-(t)-f_-(t_i)\right)\bigg\}\mathbf{d}\z \ .
\end{eqnarray}
Hence, the deviations are integrable, ${\mathbf P}^a = \mathbf{d} F^a$, with $F^1=F^2=0$, and
\begin{eqnarray}
F^3&=&-\bigg\{\left(f_+(t)-f_+(t_i)\right) \int
\beta_+(\z) \, \dd \z
\nonumber\\
&{}& + \left(f_-(t)-f_-(t_i)\right) \int
\beta_-(\z) \, \dd \z \bigg\} \ .\quad
\end{eqnarray}
It is interesting that the same result was found for the RZA model by employing Hodge--de Rham theory \cite{rza4}, while it should be noted that the whole solution is not integrable due to the nonintegrability of the initial metric, allowing for nonvanishing curvature and curvature evolution.
%

\subsection{Cosmological lattice model}\label{SubSec:CosmoLattice}

The fulfillment of Eq.~\eqref{Eq:TrueIntConst} in Proposition~\ref{Prop:TrueIntConst} can be ensured by a proper choice of the initial conditions. Once the integral constraint is satisfied at some initial time, it propagates throughout the evolution, guaranteeing the conservation of the total mass, $M_{\cD_H}\equiv \averageH{\varrho} a_{\cD_H}^3$, and the Yamabe functional, $Y_{\cD_H}\equiv\averageH{\mathcal R}a_{\cD_H}^2$,  on a certain scale of homogeneity $L_H$, while matching the corresponding FLRW evolution.

Referring back to the Szekeres variables introduced in Sec.~\ref{Sec:SzekeresModelsintro}, we find that~\eqref{Eq:TrueIntConst} reduces to the 
vanishing of the integrals of $\beta_\pm(\z)$ on $\cD_H$.
\begin{lemma}\label{Lem:VanishingIntBeta}
In Szekeres class II solutions, the necessary and sufficient condition for zero backreaction on $\cD_H$ is expressed as
\begin{equation}\label{Eq:TrueIntConstSzeVar}
\int_{\cD_H} \beta_\pm(\z) \,{\mathrm d} \z=0 \ .
\end{equation}
\end{lemma}
Appendix~\ref{AppSec:ProofOfLemmata3-4} provides the proof of Lemmata~\ref{Lem:ConsRho-3R} and~\ref{Lem:VanishingIntBeta}  at one go.
The conditions~\eqref{Eq:TrueIntConstSzeVar},~\eqref{Eq:ConsIfonlyIfClassII},~\eqref{Eqns:ConClassIILem}, and~\eqref{Eq:TrueIntConst} are equivalent, and they reduce the average model to the FLRW background on the scale of homogeneity $L_H$. Furthermore, the local requirement for the FLRW limit, $\beta_\pm(\z)=0$, finds in~\eqref{Eq:TrueIntConstSzeVar} its generalization to the Friedmann limit of the average solution. 
\begin{figure}[h]
\includegraphics[scale=0.35]{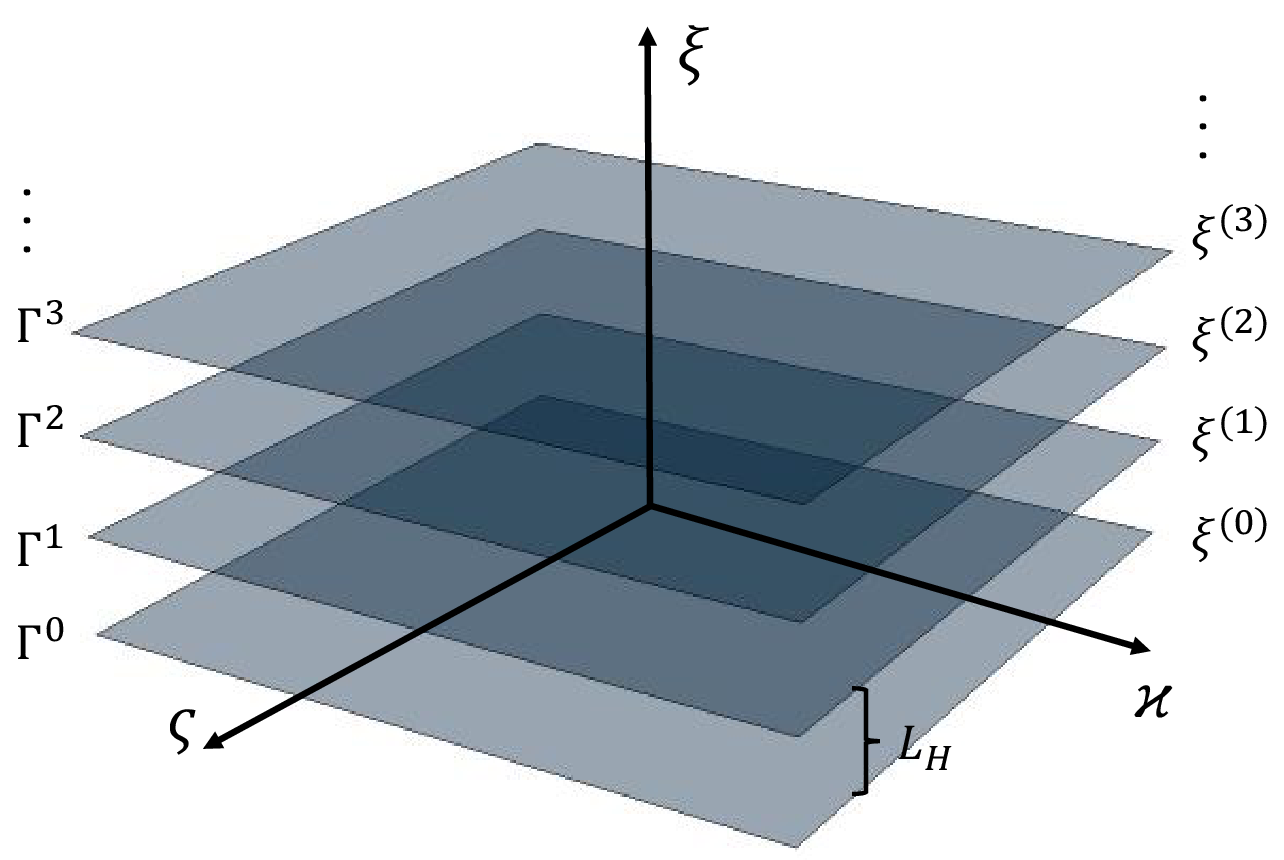}
\caption{\label{Fig:Fig-Lattice} Cosmological lattice model built from Szekeres class II solutions. 
Spatial cells comoving with a given background model, and on a scale where the integral constraint is satisfied, can be matched through $\z=$ const. surfaces. Each cell consists of a fully inhomogeneous region described by the Szekeres class II exact solution. The FLRW model emerges by construction from the spatial average of the solution on the homogeneity scale $L_H$.
}
\end{figure}

Given this integral property, we can build an idealized but exact lattice model of the Universe by matching periodic cells infinitely extended in $\x$ and $\y$ but satisfying~\eqref{Eq:TrueIntConstSzeVar} on $\left[\z^{_{(i)}},\z^{_{(i+1)}}\right]$, see Figure~\ref{Fig:Fig-Lattice}. Such cells can be matched across hypersurfaces $\Gamma^i$: $\z=\z^{_{(i)}}= \text{const.}$, described by the comoving coordinates $y^c=\left(t,\x,\y\right)$.
Denoting the  coordinates of our four-dimensional manifold by $x^\nu=\left(t,\x,\y,\z\right)$, we find the  induced metric on $\Gamma^i$ (first fundamental form)~\cite{Poisson:RelativisticToolkit}: 
\begin{eqnarray}
h_{c d}&=&g_{\mu\nu}\t{e}{\mu}{c} \t{e}{\nu}{d} \ \ (  \text{on}  \;  \;\Gamma^i: \z =\z^{_{(i)}}= \text{const.} )
\nonumber
\\
&=&\text{Diag}\left[-1, \frac{a^2}{1+\frac{k_0}{4}(\x^2+\y^2)}, \frac{a^2}{1+\frac{k_0}{4}(\x^2+\y^2)}\right] \ ,   
\nonumber
\\
&{}&\text{with} \quad \t{e}{\nu}{b}=\frac{\partial x^{\nu}}{\partial y^{b}}=\t{\delta}{\nu}{b} \ ,
\end{eqnarray}
where $b, c, d= 0, 1, 2$ and $\mu, \nu = 0, 1, 2, 3 $.
On the other hand, the extrinsic curvature (second fundamental form) identically 
vanishes on $\Gamma^i$:
\begin{equation}\label{Eq:ZeroKab}
K_{c d}=n_{\mu ; \nu} \, \t{e}{\mu}{c} \,\t{e}{\nu}{d}= 0 \ , \quad \text{(on    $\;\Gamma^i$)} \ .
\end{equation}
Above, $\mathbf{n}$ is the unitary vector normal to $\Gamma^i$,
\begin{eqnarray}
n_\nu
=a(t) \, \G \, \t{\delta}{3}{\nu} \ ,
\end{eqnarray}
where $\G$, appearing in the coframe set \eqref{coframesII}, was defined in Section~\ref{Sec:SzekeresModelsintro}.

Hence, both the first and second fundamental forms are continuous across the matching hypersurfaces ($\Gamma^i$: $\z=\z^{_{(i)}}$) as long as we take the same associated background on the whole manifold. Strictly speaking, our model comprises a set of periodic deviations matched along surfaces of constant comoving coordinates 
$\z$ of a fixed background model. 
Of particular relevance here is that such a background and the deviations altogether make up an exact solution of the Einstein equations. 
Similar models, but in the context of a generalized Szekeres class II solution with heat flow, are examined in~\cite{NajeraSussmanInPreparation}.

The cosmological lattice model introduced here has its counterpart in the common architecture of Newtonian simulations. The potential to analyze the large-scale properties of the Universe is, however, limited due to the requirement of vanishing backreaction, as in Newtonian torus-models \cite{BuchertEhlers}. However, in contrast to Swiss-cheese models, there is no need to include FLRW regions to match the inhomogeneous cells. Instead, the FLRW model emerges as an average property of the solution.  Each cell is made up of expanding and collapsing regions (undergoing pancake collapse), whose dynamics are described by the Szekeres exact solution reinterpreted in the language of RZA (which provides a straightforward connection to Newtonian models of structure formation).

\subsection{Functional evaluation and correspondence with Newtonian exact solutions}\label{SubSec:RZAandCPT}

So far, our study has been restricted to comparing the line-elements and examining their time evolution. However, a complete and consistent program should include an analysis of the functional evaluation, one of RZA's most powerful tools, where the Zel'dovich extrapolation idea~\cite{ZA}, is extended to all the relevant fields (not only the density)~\cite{Buchert1989AA,BuchertZA}. 

One interesting feature of these models  
is the identically vanishing of the second and third principal scalar invariants of the deformation matrix, $J^{(2)}$ and $J^{(3)}$ in equations~\eqref{Eq:Jacobian}-\eqref{Eq:Invariants}; then:
\bse\label{Seq:JacbFuncEval}
\begin{eqnarray}
   J&=& a^3 \left( 1 + {J}^{(1)} \right)= a^3 \left( 1 + P \right) \ ;
\\
   &{}& \Rightarrow \quad \J=J /a^3 =1+P \ ,
\end{eqnarray}
\ese
which includes only terms up to first-order in the deformation field. This exclusive appearance of the first invariant, known as ``exact body of RZA,'' is a characteristic of the anisotropic pancake collapse as a result of the dominance of the first invariant (see~\cite[appendix A]{vigneron2019dark} for a detailed discussion). Note that the second and third invariants do not vanish for a general RZA model. 

Similar results are valid for the other functionals, where the Szekeres quantities are linear in the deformation field, $\t{P}{a}{i}$, while retaining the determinant exact (linear in $\t{P}{a}{i}$ as well).  
In addition to the density $\varrho=\varrho_i J^{-1}$, the functionals of the scalar expansion and shear tensor read:
\bse
\begin{eqnarray}
\Theta&=&3 \, \frac{\dot{a}}{a}  +\frac{ \dot P }{\J} \ , 
\label{Eq:ThetaClassII}
\\
\sigma^{ i}_{\ j}&=&\frac{1}{\J}(\dot{P}^i_{\ j}-\frac{1}{3} \, \dot{P} \, \delta^i_{\ j}) \ .
\label{Eq:ShearClassII}
\end{eqnarray}
\ese
Particularly, the shear tensor simplifies to
\bse
\begin{eqnarray}\label{Eq:EqSigmaFunc}
\sigma^{ i}_{\ j}&=&\Sigma \times \hbox{Diag}\left[1,1,-2\right]\ ,
\\
\Sigma &\equiv& -\frac{1}{3} \, \frac{\dot P}{\J}  \ .
\end{eqnarray}
\ese
The expansion tensor can be reconstructed from the above expression for the expansion scalar (its trace) and the shear tensor (its antisymmetric trace-free part),   
\begin{equation}
\t{\Theta}{i}{j}
=\t{\sigma}{i}{j} \,+\,\frac{1}{3} \Theta \, \t{\delta}{i}{j}=\text{Diag}\left[\frac{\dot{a}}{a},\, \frac{\dot{a}}{a}, \, \frac{\dot{a}}{a} + \frac{\dot P}{\J}\right] \ ,
\end{equation}
which splits into the background and the exact deviation contributions.
As we will see throughout this section, this feature is not exclusive to the expansion tensor but all the relevant Szekeres functionals. The determinant and scalar curvature have the same form, while the trace-free variables (the trace-free part of the Ricci tensor, the shear and gravitoelectric part of the Weyl tensor) only contain the deviation contribution, since they vanish in the background. However, focusing on the deviation and examining the peculiar-expansion tensor, we find that it has only one nontrivial component,  
\begin{equation}
\t{\theta}{i}{j} \equiv \text{Diag}\left[0, \, 0, \, \dot P/\J \right] \ ,
\end{equation}
reinforcing our initial picture of a locally one-dimensional deformation in a homogeneous universe model. 
Recall that this is the description seen from the geodesic and irrotational frame coinciding with the eigenframe of the expansion tensor (and, in turn, coplanar with the gravitoelectric Weyl tensor eigenframe). 

While the gravitomagnetic part of the Weyl tensor vanishes identically, the functional of the gravitoelectric part is diagonal, and its only nonvanishing scalar reads: 
\bse
\begin{eqnarray}
\t{E}{i}{j} &=&\Psi_2 \times \hbox{Diag}\left[1,1,-2\right] \ ,
\\
\Psi_2  &\equiv& \frac{1}{  3 \J}
\left\{
   2 \frac{\dot a}{a} \dot P + \ddot P
\right\} \ .
\end{eqnarray}
\ese
We remark that this could have been the starting point of the paper: once we consider an irrotational dust fluid model with a fluid-flow foliation and a locally one-dimensional deformation field (as in this case), the above exact functionals invariantly characterize the Szekeres solutions (see the coordinate-independent definition of these solutions enunciated in Sec.~\ref{Sec:SzekeresModelsintro}).

On the other hand, the Ricci tensor and scalar curvature are given by
\begin{eqnarray}
\t{\CR}{i}{j}
=
\begin{cases} 
2\frac{k_{0}}{a^{2}} 
-\frac{1}{ \J}
\left(
3\frac{\dot a}{a}\dot{P}+\ddot P
\right)
 \ , 
 \hspace{1.1em} i=j=1, 2 , 
\\
2\frac{k_0}{a^2} 
-\frac{2}{ \J}
\left(
3\frac{\dot a}{a}\dot{P}+\ddot P
\right)
\ ,   
 \hspace{2.em} i=j=3 , 
 \\
0 \ ,  \hspace{10.5em} \hbox{otherwise,} \\  
\end{cases}
\end{eqnarray}
and
\begin{equation}\label{Eq:FuncR3d}
\CR=6\,\frac{k_0}{a^2}
-\frac{4}{\J} 
\left(
3\, \frac{ \dot{a} }{ a} \dot{P}
+  \ddot{P}
\right) \ ,
\end{equation}
from which we can determine the trace-free symmetric part of the Ricci tensor:
\bse\label{Seq:TauFuncEval}
\begin{eqnarray}
\t{\tau}{i}{j} = \t{\CR}{i}{j}-\frac{1}{3}\CR \, \t{\delta}{i}{j} &=&T \times \hbox{Diag}\left[1,1,-2\right] \ ,
\\
T &\equiv& \frac{1}{3 \, \J}
\left(3\, \frac{\dot a}{a}\dot{P}+\ddot P\right)  \ . \quad
\end{eqnarray}
\ese
All of the above functionals are exact and yield the well-known Szekeres quantities when the deformation field is substituted by its expression in terms of the Szekeres functions, see Appendix~\ref{SecApp:FuncEvaluationRZA}.  
Note that these functional expressions are not strictly linear, as could be interpreted from Kasai's discussion in~\cite{Kasai1995} since we keep the determinant exact ({\it i.e.}, we do not linearize expressions of the form $1/\J\equiv (1+P)^{-1}\not\approx 1-P$).  This fact is unexpected, since RZA arises from a perturbative analysis, and gives strong support to the functional extrapolation of the perturbed coframes, defined for all variables in \cite{rza1}.
 
The vanishing of the second and third invariants implies a local one-dimensional kinematical motion without three-dimensional symmetry. With this property, the Szekeres class II solutions are the relativistic analog of the three-dimensional class of Newtonian solutions examined in ~\cite{Buchert1989AA}. Both solutions correspond to a class of locally one-dimensional flow models and lead to the same nonlinear evolution of dust structures, governed by the following equation for the density contrast $\delta \equiv (\varrho - \varrho_b)/\varrho_b \ ; \ -1 \le \delta < \infty$:
\begin{equation}\label{Eq:deltarhoclassII}
\ddot{\delta} + 2 H \dot{\delta} - 4 \pi \varrho_b \delta -\frac{2}{1+\delta} \dot{\delta}^2 - 4 \pi \varrho_b \delta^2 = 0 \ ,
\end{equation}
which acquires a Lagrangian linear form for the variable $\Delta \equiv (\varrho - \varrho_b)/\varrho = \delta / (1+\delta)\ ;\ -\infty <\Delta<1$ \cite{BuchertVarenna,Roy:generalbackground}:
\begin{equation}\label{Delta}
\ddot{\Delta} + 2 H \dot{\Delta} - 4\pi \varrho_b \, \Delta = 0 \ .
\end{equation}
For Szekeres models, the previous equation is obtained from the Raychaudhuri equation, after using the corresponding (exact) functional evaluations~\cite{ishak2012growth,IshakPeel2012LargeScale}, see~\cite{Buchert1989AA} for details on the Newtonian solution. 

The preceding discussion can be summarized in the following theorem.
\begin{theorem}\label{Th:Class2Newt}
The class II of the Szekeres solutions forms the general-relativistic analog of the locally one-dimensional Newtonian solutions of~\cite{Buchert1989AA}, that are kinematically characterized by the vanishing of the second and third principal scalar invariants of the peculiar-expansion tensor at a FLRW background.
\end{theorem}

\begin{remark}
The Newtonian class of three-dimensional solutions without symmetries is restricted to initial data that are composed of sets of potential two-surfaces with vanishing Gaussian curvature, see~\cite{buchertgoetz,Buchert1989AA}. In applications it is difficult to find initial data that have no initial singularities beyond the trivial case of plane symmetry that corresponds to cylinders as special cases of surfaces with vanishing Gaussian curvature. In view of this it is evident that the use of this form of the solutions for generic initial data finds its natural realization in the RZA approximation, where still the first principal scalar invariant of the expansion tensor is dominating, but the higher invariants are nonvanishing. We also note that this restriction does not apply to Szekeres class I solutions, where several applications of realistic initial data are possible~\cite{hellaby1996null,Bolejko2006Struformation,bolejko2007evolution,IshakNwankwo2008,Bolejko:2009GERG,BolCelerier2010,KrasBol2011,Nwankwo_2011,BolSuss2011,IshakPeel2012LargeScale,SussBol2012,WaltersHellaby2012,Buckley2013,IshakPRL2013,Vrba:2014,sussman2015multiple,sussman2016coarse,Hellaby2017,gaspar2018black}.
\end{remark}

\section{Szekeres Class I and RZA}\label{Sec:CommentsClassI}

Turning now to the class I models, we may point out that, unlike class II, this class does not admit a decomposition in terms of separable functions, a necessary condition to express the Szekeres models in the language of RZA. 
The presence  of a general scale factor, $\s(t,\z)$, and the nonseparability of the solutions into space and time functions break with the spirit of RZA.
While the nine coframe functions can still be considered as the only dynamical field variables, their interpretation as a deformation of a global background fails. We consider this fact as a constructive hint on how the relativistic Lagrangian perturbation scheme would have to be extended. This extension points to the consideration of average properties in order to define a ``background'' that interacts with the inhomogeneities, {\it i.e.} that will include backreaction. 

Despite these observations, we will stick to the mathematical structure explained in Sec.~\ref{Sec:RZAintro} and reinterpret the field $\t{P}{a}{i}$ as a {\it local} generalization of the RZA deformation field in what follows.

For this aim, let us consider a formal coframe decomposition of the class I line-element. With the proviso of the spatial dependence of $\s$ and $f_\pm$,  the coframes \eqref{Eq:SecCofXi} can be written in a similar form to those of class II:
\begin{subequations}\label{Eqs:CofClassIIRZA}
\begin{eqnarray}
\tilde{\boldsymbol{\eta}}^{1}&=& \s e^\nu {\bf d}\x \ ;
\\
\tilde{\boldsymbol{\eta}}^{2} &=& \s  e^\nu {\bf d} \y \ ;
\\
\tilde{\boldsymbol{\eta}}^{3}&=&\s \G \W {\bf d} \z  = \s  \left(\A-\F\right) \W {\bf d} \z \nonumber
\\
 &=& \s  \left(\widetilde{\A}-\widetilde{\F}\right) \W  {\bf d} \z\ , 
\end{eqnarray}
\end{subequations}
where, as for class II, $\widetilde{\A}\equiv \A-\F_i$ and $\widetilde{\F}\equiv \F-\F_i$ with $\F_i=\F(t_i, \z)$ and Gram's matrix is obtained from~\eqref{Eq:GramsMatrix}, but here it is understood to relate to the initial metric by  $\s_{i}(\z)G_{ab}(\rv)=g_{ij}(t_{i},\rv)$,
\begin{equation}
G_{ab}=\hbox{Diag}\left[ e^{2\nu},e^{2\nu}, \widetilde{\A}^2 \W^2\right] \ .
\end{equation}
For this class, it is not in general possible to set the initial scale factor to unity (as for class II), but instead, $\s(t_i,\z)$ is a generic function of $\z$, $\s_i(\z)$.
Next, Eq.~\eqref{GammaijSze} leads to the expression of the only nonvanishing component of the ``generalized'' deformation field,
\begin{equation}\label{Eq:P33ClassI}
P^{3}_{\ 3} = -\widetilde{\F}/\widetilde{\A} \ ,
\end{equation}
which is formally equivalent to its homolog of class II. 

So far, this approach is only an ansatz that resembles RZA. 
To analyze its physical content, we can take advantage of the silent property of Szekeres models, under which each world line evolves independently of the others and is characterized by local quantities.\footnote{The silent property has been exploited to perform cosmological simulations beyond the Szekeres models~\cite{Bolejko:2018SilentUniv1,Bolejko:2018SilentUniv2,BolejkoOstrowski2019SilentUniv3}. }
Then, for any arbitrary world line (labeled by the comoving coordinates $\rv$), the conformal scale factor,
\begin{equation}
\s(t)\equiv\s(t,\z)|_{\z=\hbox{const.}} \ ,
\end{equation}
satisfies the Friedmann equation of an associated ``local background" with initial density,
\begin{equation}\label{Eq:RhobmuClassII}
4 \pi \varrho_b(t_i) = 3 \mu(\z)|_{\z=\hbox{const.}}\ ,
\end{equation}
and $k_0$. Consequently, the functions 
\begin{equation}
f_{\pm}(t)\equiv f_{\pm}(t,\z)|_{\z=\hbox{const.}} \ ,
\end{equation}
can be identified as the growing and decaying modes of structures on this ``local background''. Remarkably, the local quantities characterizing the world lines (density, scalar expansion, shear and spatial parts of the Weyl curvature) can be obtained from their respective functionals of RZA.  

This result gives rise to a corollary of Theorem~\ref{Th:Class2Newt} for the class I solutions.
\begin{corollary}
\label{corollary}
The dynamics of the Szekeres class I solutions  corresponds to a constraint superposition of nonintersecting world lines, each one being solution of the ``exact body'' ({\it i.e.} Szekeres class II) of RZA, but with different ``local backgrounds''. Consequently, all relevant quantities have the same local functional expressions as in class II.
\end{corollary}
By ``constraint superposition'' we mean that this superposition has to be globally consistent with Einstein's equations: the energy and momentum constraints have to be satisfied at some initial hypersurface and propagated in time. In this sense, all silent solutions are nonlocal. Locally, these solutions are generated from the free function $\mu(\z)$ which, for each fixed value of $\z$, sets an associated ``local background'' with its own local parameters, as well as the ``local growing and decaying modes''. {\textit{A priori}, there is no global background, but we may construct one through a suitable averaging operation.

Finally, let us notice that for this class the growth of structure is suppressed when a flat associated ``local background" is assumed, see Eq.~\eqref{ClassIBeta}. We can avoid this shortcoming by considering $k_0\neq0$, which would not necessarily be in contradiction with a flat global background. 
To understand this, we may see the GW formulation as a reparametrization of the original Szekeres solution, where (in class I) the parameter $k_0$ arises from a rescaling of an arbitrary function $k(\z)$: $k(\z)=k_0 \, \phi^2(\z)$; see Appendix~\ref{SecApp:RelBetweenPars}. Since Szekeres models admit regions with positive and negative $k(\z)$ matched by others with $k(\z)=0$, this property remains valid (although somewhat hidden) in the GW parametrization.

\subsection{Functional evaluation}

The discussion and formulas presented in Section~\ref{SubSec:RZAandCPT} apply equally to class I, with the proviso that the results are only valid for each fluid-element  world line. As was discussed above, these models can be reinterpreted as a superposition of local solutions of RZA with space-dependent or associated ``local backgrounds". 
\begin{remark}
In class I, all relevant fields have locally the same functional expressions as in class II.  
\end{remark}
The functional evaluation is carried out as in class II:
it is only necessary to replace the global scale factor $a(t)$ by the space-dependent scale factor $\s(t,\z)$ in~\eqref{Seq:JacbFuncEval}-\eqref{Seq:TauFuncEval}.  Here as well, all functionals are exact and yield the well-known Szekeres quantities when the deformation field is substituted by its expression in terms of the Szekeres metric functions, see Appendix~\ref{SecApp:FuncEvaluationRZA}. 

Notably, for this class, the density contrast satisfies an equation equivalent to~\eqref{Eq:deltarhoclassII}, but in terms of averaged quantities~\cite{ishak2012growth,IshakPeel2012LargeScale,NonsphericalSzePerts}, indicating again that averaging could play a key role in evaluating these models. 

\subsection{A note on admissible initial data for class I}\label{SubSec:AdmissibDataClassI}

As we have seen, in contrast to class II, the class I dynamics cannot be expressed in terms of a deformation field with respect to a global background. The nontrivial spatial dependence of the conformal scale factor introduces another degree of freedom, spoiling the basic architecture of Lagrangian perturbations.

To gain a better insight into this issue, let us examine the backreaction term and try to impose integral constraints on the deformation field (as we did for the class II). Using the kinematical functionals for the class I (which have the same mathematical structure as their equivalents of class II), we obtain for the backreaction functional: 
\begin{eqnarray}
\quad\frac{1}{6}\CQ_{\cD}=
\average{\left(\frac{\dot \s}{\s}\right)^2} 
- \average{\frac{\dot \s}{\s}}^2 - \average{\Sigma}^2
\nonumber
\\
+2\average{\frac{\dot \s}{\s}}\average{\Sigma} - 2 \average{\frac{\dot \s}{\s}\Sigma}  \ .
\label{Eq:QdClassI}
\end{eqnarray}
In analogy to class II, a natural choice for an integral constraint could be,
$\averageH{\Sigma}=\langle{\dot P/\J}\rangle_{\cD_H} = 0$, removing only two backreaction terms in \eqref{Eq:QdClassI} on the homogeneity scale $L_H$. 
But, this constraint does imply the vanishing of neither $\averageH{P/\J} = 0$, $\averageH{P/J} = 0$ nor $\langle\dot P / J \rangle_{\cD_H}= 0$. 
However, by inspecting the commutation rule, we can convince ourselves that none of these conditions leads to the vanishing of the remaining terms in $\CQ_{\cD_H}$.

All this suggests that we have to follow a more general approach for class I, where constraints are imposed on the entangled set of independent functions. An illustrative example comes from the parabolic LTB model (a subclass of the Szekeres class I solutions) averaged on a spherical domain $\cB$, for which  $\CQ_{\cB}=0$, but  neither $\langle\Sigma\rangle_{\cB}= 0$ nor the independent terms in~\eqref{Eq:QdClassI} trivially cancel, see Appendix~\ref{App:LTBSubcase}. 
%
\section{Concluding remarks}\label{Sec:DiscFinalRemarks}

We have thoroughly investigated the connection of the Szekeres exact solutions with relativistic Lagrangian perturbation theory in terms of its first-order member RZA. For the analysis, we restricted the RZA models to those with an irrotational dust source and a fluid-orthogonal foliation of the spacetime, necessary conditions to be compatible with the standard formulations of the Szekeres irrotational dust solutions. In particular, we employed a formulation of these solutions due to Goode and Wainwright, where the metric is written in terms of an associated background and exact deviations thereof. 

Within this framework, we found that the class II solution is exactly contained in RZA as a particular case known as ``exact body'', where the second and third principal scalar invariants of the deformation matrix identically vanish, and the dynamics is characterized by the exclusive appearance of the first invariant.  A similar result holds for the 
peculiar-expansion tensor. Remarkably, this class constitutes the general-relativistic analog of the locally one-dimensional Newtonian solutions introduced in~\cite{Buchert1989AA}, which are likewise characterized by the vanishing of the second and third principal scalar invariants of the peculiar-expansion tensor. 

All Szekeres relevant fields are reproduced by the exact body functionals of RZA. This reinforces our initial assertion that the RZA functional evaluation, an extension of the Zel'dovich extrapolation idea to any dynamical quantity, is more than a mere perturbative evaluation. The nonlinearities encoded in the functional definitions allow to cover exact nonlinear Szekeres solutions. 

For class II, the integral constraint $\averageH{P/\J}=0$ is a necessary and sufficient condition for the vanishing of kinematical backreaction on a domain of homogeneity $\cD_H$, providing sense to a physical background solution and exact fluctuations.
This constraint can be traced back to the conservation laws for  the FLRW density and curvature, which are also properties of the inhomogeneous class II model on $\cD_H$. The presence of the backreaction term \eqref{offsetbackreaction},  is thus, for this class, a consequence of an offset in the model initial data, which in the context of the early Universe requires fine-tuning of initial conditions.
It turns out that spatial cells satisfying the integral constraint can be smoothly matched across surfaces of constant comoving coordinates. Proceeding along these lines, we obtained a cosmological lattice model that mimics the Newtonian periodicity conditions, where the average of periodic (exact) fluctuations cancels at a certain scale of inhomogeneity ($L_H$), giving place to an isotropic and homogeneous background solution.

These assertions are not valid for class I, which has a more complex mathematical structure. Despite this more general property, the global solution can be thought of as a superposition of world lines, where each one obeys the RZA model equations. For a clear understanding of this interpretation, it is essential to bear in mind that Szekeres models belong to the family of silent solutions of general relativity, where the evolution of each world line is local. Since fluid lines are decoupled from each other, one may look at the solution as a set of independent world lines, which globally satisfy the Einstein equations' constraints. In this sense, we have paraphrased the `silence property' in the language of RZA. Consequently, all relevant fields have the same functional expressions as in class II, their evaluation is strictly local, and it is carried out along each world line.

We highlighted in several places that spatial averaging may provide the key to construct an effective background and deviations thereof also for class I. Such a ``background'' is then expected to interact with the local structure formation along the lines of the investigation in \cite{Roy:generalbackground}. If successful, such a construction would also provide clues on how to generalize the Lagrangian perturbation solutions with the aim to also include Szekeres class I solutions as well as their exact averages.

Finally, let us summarize in a theorem the most relevant result of the article. 
\begin{theorem}
For a suitable set of initial conditions, the exact body subcase of RZA corresponds to the Szekeres class II exact solution of the Einstein equations.
\end{theorem}
The proof of the theorem is implicit throughout Section~\ref{Sec:ClassIIRZA}, and by suitable initial conditions we mean the initial values of Szekeres functionals. This result goes further than a simple reformulation of the Szekeres class II solution in the RZA language, where the nontrivial evolution is represented by a single dynamical variable, {\it i.e.}, the deformation field. 
It strengthens RZA, as formulated in~\cite{rza1,rza2,rza3,rza4,rza5}, as the correct generalization of the Newtonian Zel'dovich Approximation to relativistic cosmology.

Possible implications point to two main directions. First, RZA provides a consistent framework to generalize the Szekeres class II exact solution to more general cosmologies, namely the inclusion of a nontrivial Weyl magnetic part~\cite{rza4} and to more general fluids~\cite{rza5}. On the other hand, the known exact generalizations of the Szekeres solution can provide clues on the extension of RZA  to more general scenarios containing heat-flow~\cite{Goode1986HeatFlow,LimaMaia1985HeatFlow,Baysal1992Heatflow,NajeraSussmanInPreparation}, viscosity~\cite{TomimuraMotta1990aViscosity,MottaTomimura1990bViscosity}, and electromagnetic fields~\cite{TomimuraWaga1987EMfields,LimaNobre1989EMfields} (see~\cite{kras1} for a summary of the generalizations of Szekeres solutions). 

Second, the structure of Szekeres class I solutions, as elucidated in Corollary~\ref{corollary},
paves the way of a possible strategy to arrive at a more general form of a Lagrangian structure formation theory. We have demonstrated that the tools developed for the spatially averaged RZA in \cite{rza2} can be directly employed to perform an averaging operation of a structure formation model corresponding to Szekeres class I solutions, where the ``local background'' has to be included into the averaging process. It is to be expected that such a model includes backreaction that in turn leads to an interaction between structure and the evolving average model. Through a construction via the structure of Szekeres class I solutions, such a model and its correspondent approximation would contain nontrivial exact solutions. {\it e.g.} the averaged LTB model with curvature. \\

{\bf Acknowledgments:}
This work is part of a project that has received funding from the European Research Council (ERC)
under the European Union's Horizon 2020 research and innovation program (grant agreement ERC advanced grant 740021-ARTHUS, PI: TB).  The work of IDG is supported by the DGAPA-UNAM postdoctoral grants program; also IDG acknowledges support from Grants SEP-CONACYT 282569, DGAPA-UNAM (IN112019) and CONACYT CB-2014-01 No. 240512. IDG also acknowledges hospitality during a working visit at CRAL-ENS, Lyon.
We wish to thank Juan Carlos Hidalgo, Sergio Mendoza, 
Jan Ostrowski and Roberto Sussman for helpful discussions.

\begin{appendix}

\section{Relation between Szafron and Goode-Wainwright parametrizations}\label{SecApp:RelBetweenPars}

This appendix shows the relation between the GW and Szekeres-Szafron parametrizations with $\kappa p(t)=-\Lambda$. 
Our presentation generalizes the one followed by Krasi\'nski ([Sect.~2.5]~\cite{kras1}) and Pleba\'nski and Krasi\'nski ([Sect. 19.8]~\cite{kras2}), whose modifications account for the inclusion of the cosmological constant. This fills a gap in the literature, enhancing the GW formulation of class I to include $\Lambda\neq0$. For Class II, such a generalization is due to Meures and Bruni~\cite{SzeLamddanot0Meures}.

The plan is to reproduce the steps in~\cite{kras1,kras2} to relate the arbitrary functions of one parametrization to the arbitrary functions of the other, and then confirm that the scale factor $e^\alpha$ in Eq.~\eqref{Eq:GeneralSzeMetric} can be written as
\begin{equation}\label{EqApp:DefF-a}
e^\alpha=\W \s \G\equiv \W \s \left(\A-\F\right) \ .
\end{equation}
But, instead of showing by direct substitution that the parametric expressions for $\F$, valid only for $\Lambda=0$, satisfy~\eqref{Eq:Ftt}, we will verify that $\F$, as defined in~\eqref{EqApp:DefF-a}, 
can be expressed as a linear combination of the growing and decaying modes on a general background~\cite{PeacockBookCosPhys}:
\begin{align}
f_{\pm} \propto
\begin{cases}
\left(\dot{\s}/\s\right) \int_0^{\s} \dot{\hs}^{-3} \dd \hs \ , \quad &\text{growing mode:} \;  f_+ \ ,
\\
\left(\dot{\s}/\s\right), & \text{decaying mode:} \; f_- \ .
\end{cases}
\label{AppEq:G-DModesPeacock}
\end{align}
In our case, $\s$ denotes the scale factor of the ``associated background'', interpreted as ``local'' when the spatial dependence of class I is considered (the terms ``associated" and ``local backgrounds'' were introduced in the main text, see Sect.~\ref{Sec:CommentsClassI}). Hence, $f_{\pm}$ are the growing and decaying solutions of 
\begin{equation}\label{AppEq:Ftt}
\ddot{f} + 2\frac{\dot{\s}}{\s} {\dot f} -\frac{3\mu}{\s^3} f = 0 \ ,
\end{equation}
where $\s$ obeys Eq.~\eqref{Eq:FriedmannLikeEqn}. 

\subsection{The $\beta_{,\z}\neq0$ subfamily}

In the Szekeres-Szafron parametrization, the line-element reads~\cite{kras1,kras2}:
\bse\label{metricCoeffParamI}
\begin{eqnarray}
\dd s^2&=&-\dd t^2+e^{2 \alpha} \dd \z^{2} + e^{2 \beta} \left(\dd \x^{2}+\dd \y^{2}\right) \ ;
\\
e^\beta&=&\Phi(t,\z) e^{\hat{\nu} (\rv)} \ ; \label{AppEq:ebetaClassII}
\\
e^\alpha&=& h(\z) \Phi(t,\z) \beta_{,\z} 
\equiv h(\z) \left(\Phi_{,\z} +\Phi \hat{\nu}_{,\z} \right) ; \qquad
\\
e^{-\hat{\nu}}&=& \mathcal{V}_0(\z) \left(\x^{2}+\y^{2}\right) +2 \mathcal{V}_1(\z) \x \nonumber
\\
&{}& \qquad \qquad \qquad\;\;\;\,+\,2 \mathcal{V}_2(\z) \y +  \mathcal{V}_3 (\z) \ , \label{nudefeq1}
\end{eqnarray}
\ese
where $\Phi$ satisfies a Friedmann-like differential equation, 
\begin{equation}
2 \frac{\ddot{\Phi}}{\Phi}+\frac{\dot{\Phi}^{~2}}{\Phi^2} -\Lambda +\frac{k(\z)}{\Phi^2}=0 \ ,\label{EqApp:Phi_tt} 
\end{equation}
and the following relation holds:
\begin{equation}\label{EqApp:restSza}
4 \left( \mathcal{V}_0 \, \mathcal{V}_3 - \mathcal{V}_1^{~2}-\mathcal{V}_2^{~2}\right)= \left[1/h(\z)^2+k(\z)\right]. 
\end{equation}
For this class, $k$, $ \mathcal{V}_0$,  $\mathcal{V}_1$, $\mathcal{V}_2$, and $\mathcal{V}_3$ are arbitrary functions of $\z$ to be specified in the initial data.

\subsubsection{Case $k\neq0$}

To obtain the Szekeres solutions in the GW parametrization, we normalize $k(\z)$ by introducing an auxiliary function, $\phi(\z)$,
\begin{equation}\label{EqApp:k-phi-k0}
k(\z)=k_0 \phi^2(\z) \quad \Rightarrow \quad \phi=|k|^{1/2} \ ,
\end{equation}
so that $k_0=\pm 1$. Next, the scale function $\s$ is defined as
\begin{equation}\label{EqApp:Phi-phi-S}
\Phi=\phi \, \s \ .
\end{equation}
Substituting~\eqref{EqApp:k-phi-k0} and~\eqref{EqApp:Phi-phi-S} into~\eqref{EqApp:Phi_tt}, we obtain a second-order differential equation for $\s$:
\begin{equation}\label{EqApp:Stt}
2 \ddot{\s}/\s+\dot{\s}^2/\s^2-\Lambda+k_0/\s^2=0 \ ,
\end{equation}
which can be integrated to yield Eq.~\eqref{Eq:FriedmannLikeEqn}:
\begin{equation}\label{EqApp:FridClassI}
\dot{\s}^2=-k_0 + \frac{2 \mu}{\s} +\frac{\Lambda}{3}\s^2\ .
\end{equation}
In this equation, $\mu(\z)$ is an integration constant, which in the main text was identified as the initial energy density  of the associated ``local background'', $4 \pi \varrho_b(t_i) = 3 \mu(\z)$ along the world line labeled by $\z=$const.

Next, we introduce another auxiliary function, $\zeta_s (\z)$, defined by
\begin{equation} \label{EqApp:Gs1}
\epsilon\, \zeta_s^2  \equiv  \left[1/h^2+k_0 \phi^2\right] = 4 \left( \mathcal{V}_0 \mathcal{V}_3 - \mathcal{V}_1^{~2}-\mathcal{V}_2^{~2}\right) \ , 
\end{equation}
with $\epsilon=0, \pm 1$.
Then,
\begin{equation} \label{EqApp:Gs2}
   \zeta_s = \left\{ 
  \begin{array}{l l}
  |1/h^2(\z)+k(\z)|^{1/2} ,  & \hbox{if} \quad 1/h^2(\z)+k(\z) \neq 0 \ ,\\
  \zeta_s \neq 0 \;  \hbox{and arbitrary} , & \quad \hbox{otherwise.}\\
  
  \end{array} \right.
\end{equation} 
The functions $c_i$, $f$ and $\W$ arise from the reparametrizations
\begin{eqnarray} \label{EqApp:Def_ci}
\left( c_0, c_1, c_2, c_3, f \right)  = \left( \mathcal{V}_0, \mathcal{V}_1, \mathcal{V}_2, \mathcal{V}_3, \phi \right)\zeta_s^{-1} \ .
\end{eqnarray}
From~\eqref{EqApp:Def_ci} and~\eqref{EqApp:Gs1}, it is evident that Eq.~\eqref{SubEq:Rel-ci} in the text holds:
\begin{equation}
c_0 c_3-c_1^2-c_2^2=\epsilon/4 \ .
\end{equation}
The metric function $e^{-\hat{\nu}}$,  in~\eqref{metricCoeffParamI}, takes the following form in the new variables:
\begin{eqnarray}\label{AppEq:eTohatnu-a}
e^{-\hat{\nu}}&=&\zeta_s \left[c_0 (\x^2+\y^2) + 2c_1 \x+2c_2 \y+c_3 \right] \nonumber
\\
&=&\frac{\phi}{f} \left[c_0 (\x^2+\y^2) + 2c_1 \x+2c_2 \y+c_3 \right]  \nonumber
\\
&\equiv& \phi \, e^{-\nu} \ ,
\end{eqnarray}
where we have defined (as in Eq.~\eqref{Eq:eToNuGW}),
\begin{equation}
e^{\nu}\equiv f \left[c_0 (\x^2+\y^2) + 2c_1 \x+2c_2 \y+c_3 \right]^{-1} \ .
\end{equation}
Next, the function  $\W$ is defined as
\begin{equation}\label{AppEq:W=hG}
 \W = h \, \zeta_s \ .
\end{equation}
To obtain its final form, Eq.~\eqref{Eq:AandWclassIGW}, note that the first equality in Eq.~\eqref{EqApp:Gs1} implies that 
\begin{equation}
h^2=\left(\epsilon\,\zeta_s^2  - k_0 \phi^2\right)^{-1} \ ;
\end{equation}
then,
\begin{equation}
\W^2\equiv  h^2 \zeta_s^2
=\frac{\zeta_s^2}{\epsilon \zeta_s^2  - k_0 \phi^2} =\frac{1}{\epsilon  - k_0 \left(\phi/\zeta_s\right)^2}=\frac{1}{\epsilon  - k_0 f^2} \ . \qquad\quad
\end{equation}
Using~\eqref{EqApp:Phi-phi-S} and~\eqref{AppEq:eTohatnu-a}, we express the metric function $e^\beta$ in~\eqref{AppEq:ebetaClassII}  in terms of the GW variables,
\begin{equation}
e^\beta= \Phi e^{\hat{\nu}}=\left(\phi \, \s\right) \left(\phi^{-1} \, e^{\nu}\right)=e^{\nu} \s \ .
\end{equation}
On the other hand, we need~\eqref{EqApp:Def_ci} ($\Rightarrow f=\phi \, \zeta_s^{-1}$) and~\eqref{AppEq:W=hG} to find $e^\alpha$ in the GW parametrization:
\begin{eqnarray}
e^\alpha &=& h \left(\Phi_{,\z} +\Phi \hat{\nu}_{,\z} \right) =h \, \phi \left(S_{,\z} +S \nu_{,\z} \right) \nonumber
\\
&=& \left(\frac{\W}{\zeta_s}\right) \, \phi \left(S_{,\z} +S \nu_{,\z} \right) 
= \W \left(\frac{\phi}{\zeta_s}\right) \,  \left(S_{,\z} +S \nu_{,\z} \right) \nonumber
\\
&=& \W f \left(S_{,\z} +S \nu_{,\z} \right)=\s \W \left( f \frac{\s_{,\z}}{\s}+f \nu_{,\z} \right)  \nonumber
\\
&=&\s \W  \G \ .
\end{eqnarray}
In the last term of the above equation we have defined the metric function $\G$,
\begin{equation}\label{EqApp:DefFeq1}
\G\equiv f \frac{\s_{,\z}}{\s}+f \nu_{,\z}\ ,
\end{equation}
which splits into a time-independent and a time-dependent part. Let us focus on the time-dependent part:  $\s_{,\z}/\s$; $f(\z)$ will be considered at the end.

First, we note that~\eqref{EqApp:FridClassI} has a formal integral of the form,
\begin{equation}\label{AppEq:IntSParm}
\int^{\s}_0\frac{\dd \hs}{\left(-k_0 + \frac{2 \mu}{\hs} +\frac{\Lambda}{3}\hs^2\right)^{1/2}}=t-\T(\z) \ ,
\end{equation}
where $\T(\z)$ is the time of the initial singularity, and
\begin{equation}\label{AppEq:dotSParm}
\dot{\s}=\left(-k_0 + \frac{2 \mu}{\s} +\frac{\Lambda}{3}\s^2\right)^{1/2} \ .
\end{equation}
To find an equation for $\s_{,\z}$, let us differentiate~\eqref{AppEq:IntSParm} and use~\eqref{AppEq:dotSParm} to rewrite some terms conveniently,
\begin{equation}\label{AppEq:DS-Int-a}
\frac{\s_{,\z}}{\dot{\s}}-\mu_{,\z} \int^{\s}_0\frac{\dd \hs}{\hs (\dot{\hs})^3}=-\T_{,\z} \ .
\end{equation}
After some algebra and using the Friedmann-like equation for $\dot \s$ multiple times, the integral in the equation above results in
\begin{equation}\label{AppEq:Int-SdotSto3}
\int^{\s}_0\frac{\dd \hs}{\hs (\dot{\hs})^3}=\frac{1}{3 \mu} \left\{ k_0 \int^{\s}_0\frac{\dd \hs}{ (\dot{\hs})^3}+ \left(\frac{\dot{\s}}{\s}\right)^{-1}\right\} \ .
\end{equation}
This expression allows rewriting~\eqref{AppEq:DS-Int-a} as follows:
\begin{equation}\label{AppEq:}
\frac{\s_{,\z}}{\dot{\s}}=-\T_{,\z} + \frac{\mu_{,\z}}{3 \mu} \left\{ k_0 \int^{\s}_0\frac{\dd \hs}{ (\dot{\hs})^3}+ \left(\frac{\dot{\s}}{\s}\right)^{-1}\right\} \ ; \quad
\end{equation}
hence, we obtain:
\begin{equation}
\frac{\s_{,\z}}{\s}=
-\T_{,\z} \left(\frac{\dot{\s}}{\s}\right)+ 
k_0 \frac{\mu_{,\z}}{3 \mu} \left(\frac{\dot{\s}}{\s} \int^{\s}_0\frac{\dd \hs}{ (\dot{\hs})^3}\right)+ \frac{\mu_{,\z}}{3 \mu}  \ , 
\end{equation}
where the terms in parentheses are the growing and decaying solutions of~\eqref{AppEq:Ftt} (and~\eqref{Eq:Ftt}). Then, in~\eqref{EqApp:DefFeq1}, $\G$ is given by
\begin{eqnarray}
\G&=&f \frac{\s_{,\z}}{\s}+f \nu_{,\z} \nonumber
\\
&=&   \left(f \frac{ k_0 \mu_{,\z}}{3 \mu}\right) f_+ -\left(f \T_{,\z}\right) f_- + f \frac{ \mu_{,\z}}{3 \mu} +f \nu_{,\z} 
\nonumber
\\
&=&
\left( -k_0 \left( -f \frac{k_0 \mu_{,\z}}{3 \mu} \right)+f \nu_{,\z} \right) \nonumber
\\
&{}&
 - \left(\left(-f \frac{ k_0 \mu_{,\z}}{3 \mu}\right) f_++\left(f \T_{,\z}\right) f_-\right) \, ,
\end{eqnarray}
which can be rewritten as
\begin{equation}\label{AppEq:AFParenth}
\G= \left(f \nu_{,\z} -k_0 \beta_+  \right)-\left( \beta_+ f_+ +\beta_- f_-\right) \ . 
\end{equation}
Above, we have introduced the functions
\begin{equation}\label{AppEq:betaDef}
\beta_+\equiv-f \frac{ k_0 \mu_{,\z}}{3 \mu} \ ; \qquad \beta_-\equiv f \T_{,\z} \ .
\end{equation}
The first term of Eq.~\eqref{AppEq:AFParenth} corresponds to the time-independent part of $\G$ in~\eqref{EqApp:DefFeq1},
\begin{equation}\label{AppEq:AdefClasI}
\A \equiv f \nu_{,\z} -k_0 \beta_+  \ ,
\end{equation}
proving the first equation in~\eqref{Eq:AandWclassIGW}, while the second term of Eq.~\eqref{AppEq:AFParenth} is the time-dependent part of $\G$ and satisfies~\eqref{Eq:Ftt} (and~\eqref{AppEq:Ftt}) for any ``local background",
\begin{equation}\label{AppEq:FdefClasIBetas}
\F \equiv \beta_+ f_+ +\beta_- f_- \ .
\end{equation}
Some comments are in order. The solutions of Eq.~\eqref{Eq:Ftt} (or~\eqref{AppEq:Ftt})  are undetermined by a multiplicative function that is constant in time (a function of $\z$ in our case). Goode and Wainwright took advantage of this freedom to simplify the parametric solutions, and chose that constant to be proportional to $\mu$. Consequently, they defined the $\beta_-$ function as $\beta_-=f  \mathcal{T}_{,\z}/(6\mu)$. We have omitted the denominator since we are not interested in analyzing the parametric solution, and its presence is not necessary. The physics of these equations is contained in the whole term $\A(\rv)-\F(t,\z)$.

\subsubsection{Case $k=0$}

Things are much simpler when $k(\z)=0$. First, in this case we cannot use Eq.~\eqref{EqApp:k-phi-k0} to define  $\phi$ or $f$. Thus, we take
\begin{equation}
\phi\equiv\mu^{1/3}  .
\end{equation}
From~\eqref{EqApp:Phi-phi-S},
\begin{equation}
\Phi=\mu^{1/3} \s \ ,
\end{equation}
we find that $\s$ satisfies~\eqref{Eq:FriedmannLikeEqn}  with $k=0$ and $\mu=1$. %
Consequently, for $k=0$ we can take $\mu=\text{const.}$ in the Friedmann-like equations, and then use the arbitrary function $\phi$ to set the parametrization. 
Proceeding along these lines, the rest of the equations remains the same as in the previous paragraph. We only would like to highlight that, as defined in~\eqref{EqApp:DefFeq1}, the solutions of $\F$ only contain the contribution of $\beta_-$, so that we can assume $\beta_+=0$.
Note that our definitions of $\beta_\pm$ in the previous class, Eq.~\eqref{AppEq:betaDef}, are compatible with the results for $k_0=0$. 
Hence, the parametrization~\eqref{AppEq:betaDef} is valid for the whole class I. 

Let us illustrate this point with some equations. To write out Eq.~\eqref{EqApp:DefFeq1}, we need to differentiate~\eqref{AppEq:IntSParm}; for 
$\mu=\text{const.}$, Eq.~\eqref{AppEq:DS-Int-a} reduces to
\begin{equation}
\s_{,\z}/\dot{\s}=-\T_{,\z} \ .
\end{equation}
And, from the definition of $\G$, Eq.~\eqref{EqApp:DefFeq1}, we have 
 \begin{eqnarray}
\G&=&f \frac{\s_{,\z}}{\s}+f \nu_{,\z}=-f \T_{,\z} \left(\dot{\s}/\s\right) + f \nu_{,\z} \nonumber
\\
&=&f \nu_{,\z} -\beta_-f_- \ .
\end{eqnarray}
Hence, for $k_0=0$, we have 
\begin{equation}
\A= f \nu_{,\z} \ ; \quad \F=\beta_-f_- \ ; \quad  \beta_-= f \T_{,\z} \ ; \quad  \beta_+=0 \ .
\end{equation}

\subsection{The $\beta_{,\z}=0$ subfamily}

The solution for this family is given by~\cite{kras1,kras2}:
\begin{subequations}\label{metricII}
\begin{eqnarray}
\dd s^2&=&-\dd t^2+e^{2 \alpha} \dd \z^{2} + e^{2 \beta} \left(\dd \x^{2}+\dd \y^{2}\right)  \ ; 
\qquad
\\
e^{\beta}&=&\Phi(t) e^\nu \ ; 
\\
 e^\alpha&=&\Phi(t) \sigma(\x,\y,\z)+\lambda(t,\z)\ ;
\\
\sigma&=&e^\nu 
\Bigg[ \frac{1}{2} \mathcal{U}(\z) \left(\x^{2}+\y^{2}\right)+ \mathcal{V}_1(\z) \x \nonumber
\\
&{}&\qquad+ \mathcal{V}_2(\z) \y  + 2 \mathcal{Z}(\z)\Bigg]\ ;
\\
e^{-\nu}&=& 1+ \frac{1}{4} k (\x^{2}+\y^{2})\ .
\end{eqnarray}
\end{subequations}
In the above equations, $k$ is a constant, and $\Phi$ and $\lambda$ satisfy the following equations:
\bse
\begin{eqnarray}
&{}&2 \Phi \ddot \Phi+\dot{\Phi}^2-\Lambda \Phi^2+k=0 \ ;	\label{dbequalzeroaeq1}
\\
&{}&\ddot{\lambda} \Phi+\dot{\lambda} \dot{\Phi}+\lambda \ddot \Phi -\Lambda\lambda\Phi =\mathcal{U}+k \mathcal{Z}\ ,\label{dbequalzerolammdaeq1}
\end{eqnarray}
\ese
which admit integrals of the form,
\bse
\begin{eqnarray}
\dot{\Phi}^2=-k +\frac{2 \mu}{\Phi}+\frac{1}{3}\Lambda \Phi^2 \ ; \qquad\label{AppEq:FriedLikeClassII}
\\
\dot \lambda \Phi \dot \Phi +\frac{\lambda \mu}{\Phi}-\frac{\Lambda}{3} \lambda \Phi^2=\left(\mathcal{U}+k \mathcal{Z}\right)\Phi + \mathcal{X}(\z) \ ,\qquad
\label{AppEq:Lamb1Int}
\end{eqnarray}
\ese
where $\mu$ is a constant. Equation~\eqref{AppEq:Lamb1Int}  can be solved by quadrature as follows: 
\begin{equation}\label{AppEq:Quadraturelambda}
\lambda=\dot{\Phi}\left(
\int \frac{ \left(\mathcal{U}+k \mathcal{Z}\right)\Phi + \mathcal{X}}{\Phi \dot{\Phi}^3}\dd \Phi + \mathcal{Y}(\z)
 \right)
  \ .
\end{equation}
Here, $\dot{\Phi}$ denotes the square root of the right-hand side of~\eqref{AppEq:FriedLikeClassII}.

Finally, note that when $k\neq0$, it can be set to $\pm1$ by rescaling the coordinates and the arbitrary functions. Below, we will replace $k$ by $k_0$.

\subsubsection{Case $k=k_0=\pm1$}

For class II models with $k_0\neq0$, it is always possible to drop the right-hand side term of~\eqref{dbequalzerolammdaeq1} by
redefining $\lambda$, $\mathcal{U}$, and $\mathcal{Z}$,  which will be assumed throughout this section,
\begin{equation}\label{AppEq:UkZeq0}
\mathcal{U}+ k_0 \mathcal{Z}=0 \ .
\end{equation}
In the GW formulation the metric functions are reparametrized as follows:
\bse
\begin{eqnarray}
\Phi=\s(t)\ ; \qquad e^\beta&=&S e^\nu \ ;
\\
\mathcal{U}=-k_0 c_0/2 \ ;   \qquad \mathcal{Z}&=&c_0/2 \ ; 
\\
\mathcal{V}_1=c_1 \ ;  \qquad \mathcal{V}_2&=&c_2 \ ,
\end{eqnarray}
\ese
which implies
\begin{equation}
\dot{\s}^2=-k_0 +\frac{2 \mu}{\s}+\frac{1}{3}\Lambda \s^2  \ ,
\end{equation}
and
\begin{equation}
\sigma=e^\nu
\Bigg[ c_0\left(1- \frac{k_0}{4} \left(\x^{2}+\y^{2}\right)\right)+ c_1 \x + c_2 \y \Bigg] \ .
\end{equation}
To reparametrize the remaining metric coefficient, let us proceed as in the previous class,
\begin{equation}
e^\alpha=\s \sigma+\lambda= \s\left(\frac{\lambda}{\s}+\sigma\right)\equiv \s \G\ . 
\end{equation}
Considering the formal solution for $\lambda$, Eq.~\eqref{AppEq:Quadraturelambda}, under the assumption~\eqref{AppEq:UkZeq0}:
\begin{eqnarray}
\frac{\lambda}{\s}&=&\frac{\dot{\s}}{\s}\left(\mathcal{X} \int_0^\s  \frac{ \dd \hs }{\hs \dot{\hs}^3} + \mathcal{Y} \right) 
\nonumber\\
&=&\frac{\dot{\s}}{\s} \left\{\frac{\mathcal{X} }{3 \mu} \left[ k_0 \int^{\s}_0\frac{\dd \hs}{ (\dot{\hs})^3}+ \left(\frac{\dot{\s}}{\s}\right)^{-1}\right]+ \mathcal{Y} \right\}  \nonumber
\\
\label{AppEq:lclass2Int}
\nonumber\\
&=&  k_0 \frac{\mathcal{X} }{3 \mu}  \left(\frac{\dot{\s}}{\s} \int^{\s}_0\frac{\dd \hs}{ (\dot{\hs})^3}\right) +\mathcal{Y} \left(\frac{\dot{\s}}{\s}\right)+\frac{\mathcal{X} }{3 \mu} \nonumber
\\
&=&  k_0 \frac{\mathcal{X} }{3 \mu}  f_+  + \mathcal{Y} f_- +\frac{\mathcal{X} }{3 \mu} \ .
\end{eqnarray}
Above, we used~\eqref{AppEq:Int-SdotSto3} to rewrite the integral in the third line, and~\eqref{AppEq:G-DModesPeacock} to substitute the growing and decaying modes by $f_\pm$.

As for class I, we split $\G\equiv\A-\F$ into its time-dependent ($\F$) and time-independent ($\A$) parts,
\begin{eqnarray}
\G&=&\frac{\lambda}{\s}+\sigma \nonumber
\\
&=&\sigma - k_0\left( -k_0 \frac{\mathcal{X} }{3 \mu}\right) -\left(\left(-k_0 \frac{\mathcal{X} }{3 \mu}\right)  f_+ + \left(- \mathcal{Y}\right) f_- \right)  
\nonumber
\\
&=& \left(\sigma - k_0 \beta_+ \right)-\left(\beta_+ f_+ + \beta_- f_-\right) \ .
\end{eqnarray}
In the last line we can identify $\A=\sigma - k_0 \beta_+$, which coincides with the first equation in~\eqref{Eq:DefAclassII}, and $\F=\beta_+ f_+ + \beta_- f_-$, solution of~\eqref{Eq:Ftt}. 
Since $\mathcal{X}(\z)$ and $\mathcal{Y}(\z)$ are arbitrary functions of $\z$, $\beta_+\equiv-k_0 \mathcal{X}/(3 \mu)$ and $\beta_-\equiv- \mathcal{Y}$ are arbitrary   as well. In contrast to class I, $f_+$ and $f_-$ are only time-dependent functions ($\mu$ is constant for this class).

\subsubsection{Case $k=k_0=0$}

When $k_0=0$, $e^\nu=1$, $\Phi=\s(t)$, $e^\beta=\s$, and 
\begin{eqnarray}
e^\alpha&=&\s \left(  \frac{\lambda}{\s} + \sigma \right) \equiv \s \G\ ,
\end{eqnarray}
the term $\lambda/\s$ can be rewritten in the following form: 
\begin{eqnarray}
\frac{\lambda}{\s}&=&
\frac{\dot{\s}}{\s}\left(
\int_0^{\s} \frac{ \mathcal{U}\hs + \mathcal{X}}{\hs \dot{\hs}^3}\dd \hs + \mathcal{Y}(\z)
 \right) \nonumber
 \\
 &=&
\mathcal{U}  \left(\frac{\dot{\s}}{\s}\int_0^{\s} \frac{ \dd \hs}{ \dot{\hs}^3}  \right) + \mathcal{Y}(\z)\left(\frac{\dot{\s}}{\s}\right)+\frac{\mathcal{X}}{3\mu} \nonumber
\\
 &=&
 \mathcal{U}  f_+ + \mathcal{Y} f_-+\frac{\mathcal{X}}{3\mu} \ .
\end{eqnarray}
Defining $\beta_+=- \mathcal{U}$ and $\beta_-=-\mathcal{Y}$, we can verify that $\G=\A-\F$ with $\F=\beta_+ f_+ + \beta_- f_-$ and
$\A=\sigma + \mathcal{X}/(3\mu)$.
The final form of $\A$, Eq.~\eqref{Eq:DefAclassII}, follows from the parametrizations $\mathcal{V}_1=c_1$, $\mathcal{V}_2=c_2$  and $\mathcal{X}/(3\mu)+2\mathcal{Z}=c_0$.
As in the previous case, the functions $\beta_+$ and $\beta_-$ (and $c_0$) inherit the arbitrariness from $\mathcal{X}$ and $\mathcal{Y}$, and $\F$ is a solution of~\eqref{Eq:Ftt} with $f_{\pm}=f_{\pm}(t)$.

\section{Parametric solutions for the $\Lambda=0$ case}\label{App:ParametricSols}

For completeness, we show the parametric solutions found by Goode and Wainwright in~\cite{GW1}, considering $\Lambda=0$.
As we mentioned in Sect.~\ref{GWpara} and discussed in Appendix~\ref{SecApp:RelBetweenPars}, our definition of $\beta_-$ differs from the one introduced in~\cite{GW2}.  The GW solution (valid for both classes) is given by~\cite{GW1,kras1,kras2}:
\begin{subequations}
\begin{equation} \label{Eq:Sparam} 
\s=\mu \frac{dh(\tau)}{d\tau} \ ; \qquad 
t-\mathcal{T}(\z)=\mu h(\tau) \ ,
\end{equation} 
with
\begin{equation} \label{Eq:h_n}
  h(\tau) = 
  \begin{cases}
    \tau-\sin\tau,  & \quad k_0=+1 \ , \\
    \sinh\tau-\tau, & \quad k_0=-1 \ , \\
    \tau^3/6,       & \quad k_0=0\ .\\
  \end{cases} 
\end{equation}
\end{subequations}
For $\Lambda=0$, the solutions of~\eqref{Eq:Ftt} can be cast into the form:
\begin{subequations}  \label{Eq:fs}
\begin{equation} \label{Eq:f+}
  f_{+} = 
  \begin{cases}
    (6\mu/\s)\,[1-(\tau/2)\cot(\tau/2)]-1, & \quad k_0=+1 \ , \qquad\\
    (6\mu/\s)\,[1-(\tau/2)\coth(\tau/2)]+1, & \quad k_0=-1 \ ,\\
    \tau^2/10, & \quad k_0=0 \ ,\\
  \end{cases} 
\end{equation}
and
\begin{equation} \label{Eq:f-}
  f_{-} = 
  \begin{cases}
    (6\mu/\s)\,\cot(\tau/2), & \quad k_0=+1 \ , \\
    (6\mu/\s)\,\coth(\tau/2), & \quad k_0=-1 \ ,\\
    24/\tau^3, & \quad k_0=0 \ .\\
  \end{cases} 
\end{equation} 
\end{subequations} 
%
%

\section{FLRW limit and Cartesian coordinates}\label{App:FLRW-Cartesians}

When $\beta_{+}=\beta_{-}=0$, the FLRW limit emerges in the ``Goode and Wainwright representation of the FLRW models''~\cite{GW1,kras1,kras2}. 
\\

\noindent
For class II: 
\begin{eqnarray}
\dd s^2&=&-\dd t^2+a(t)^2\left[\bar{\A}^2 \dd \z^{2} + e^{2\nu} (\dd \x^{2}+ \dd \y^{2})\right];\nonumber
\\
e^{\nu}&=&\left[1+\frac{1}{4}k_0 \left(\x^{2}+\y^{2}\right) \right]^{-1} \ ; \nonumber
\\
\bar{\A}&=&e^{\nu}\Bigg\{c_0(\z) \left[1-\frac{1}{4}k_0 \left(\x^{2}+\y^{2}\right)\right] \nonumber
\\
&{}&\qquad \qquad \quad \quad \;+ c_1(\z) \x +c_2(\z) \y\Bigg\}\ . \label{Eq:FLRWClassII}
\end{eqnarray}
Above, we have substituted the scale factor $\s(t)$ by the FLRW function $a(t)$. Since~\eqref{Eq:FriedmannLikeEqn}
is nothing more than the Friedmann equation, its solution is the FLRW scale factor.

The choice $c_0=c_2=0$ and $c_1=1$ yields~\cite{kras1,kras2}:
\begin{equation}\label{Eq:FriedEqIClassI}
\dd s^2=-\dd t^2+\frac{a^2(t)}{\left[1+\frac{1}{4} k_0 \left( \x^2+ \y^{2} \right) \right]^2}(\x^{2} \dd \z^{2}+\dd \x^{2}+ \dd \y^{2}). 
\end{equation}
The transformation
\begin{equation}\label{App:TransToCartII}
\x=r=\sqrt{x^{2} + y^{2}}, \quad \y=z,\quad \z=\varphi=\arctan \left(\frac{y}{x}\right), 
\end{equation}
takes \eqref{Eq:FriedEqIClassI} to Cartesian coordinates:
\begin{equation}
\dd s^2=-\dd t^2+\frac{a^2(t)}{\left[1+\frac{1}{4} k_0 \left( x^2+ y^2+z^2 \right) \right]^2}(\dd x^2+\dd y^2+\dd z^2)\ . \label{RW-Sze-F1}
\end{equation}
\\

 \noindent
For class I: 
\begin{equation}\label{Eq:FLRWClassI}
\dd s^2\!=\!-\dd t^2+a^2 
\Bigg[ 
 \W^2 f^2 \nu_{,\z}^{~2} \dd \z^{2}+ e^{2\nu} \left(\dd \x^{2}+ \dd \y^{2}\right)
\Bigg].
\\
\end{equation}
Setting $c_1=c_2=0$,  $c_3=4 c_0=1$, and identifying $f=\z=r$,
the $O(3)$ orbits of the FLRW class correspond to spheres of the Szekeres spacetime~\cite{kras1,kras2}:
\begin{equation}
\dd s^2=-\dd t^2+a^2
\Bigg[ 
\frac{1}{1-k_0 r^2} \dd r^2 
+ e^{2\nu} \left(\dd \x^{2}+ \dd \y^{2}\right)
\Bigg], 
\end{equation}
\begin{equation}
\hbox{where} \qquad e^{\nu} = \frac{r}{\left[\frac{1}{4} \left(\x^2+\y^2\right) +1\right]} \ .
\end{equation}
The FLRW line-element in spherical coordinates, 
\begin{equation}
\dd s^2=-\dd t^2+a^2(t)\left[\frac{\dd r^2}{1-k_0 r^2}+r^2\left(d\vartheta^2+\sin^2\vartheta d\varphi^2\right)\right], \label{RW-Sze-SpheCoor}
\end{equation}
is obtained from the transformation
\begin{equation}
\x=2\cot\left(\frac{\vartheta}{2}\right) \cos \varphi, \quad \y=2\cot\left(\frac{\vartheta}{2}\right) \sin \varphi \ .
\label{T1FLRWlimitBneq0}
\end{equation}
Then, the metric in Cartesian coordinates, Eq.~\eqref{RW-Sze-F1}, is recovered after identifying
\begin{eqnarray}
&{}&x=\tilde{r} \sin \vartheta \cos \varphi  , \;   y=\tilde{r} \sin \vartheta \sin \varphi  , \;  z=\tilde{r} \cos \vartheta \ , 
\qquad\;
\label{T2FLRWlimitBneq0}
\\
&{}&\hbox{with} \qquad \tilde{r} =\left(1-\sqrt{1-k_0 r^2}\right) / \left({ k_0 r/2}\right) \ .  \label{RFLRWlimitBneq0}
\end{eqnarray} 
The successive change of variables~\eqref{T1FLRWlimitBneq0}-\eqref{T2FLRWlimitBneq0}  
can be summarized in one transformation:
\begin{equation}\label{App:TransToCartI}
\x=2 x \frac{\chi+z}{\chi_{xy}^2}, \quad \y=2 y \frac{\chi+z}{\chi_{xy}^2}, \quad \z=\frac{\chi}{1+\frac{1}{4} k_0 \chi^2} \ ,
\end{equation}
where we have used $\chi$ and $\chi_{xy}$ as shorthands for $\sqrt{x^2+y^2+z^2}$ and $\sqrt{x^2+y^2}$.


\section{Spatially averaged equations}\label{App:BackAveEqns}

The kinematic expansion and acceleration laws of an FLRW spacetime find their generalization in a general, neither homogeneous nor isotropic, spacetime through exact volume-expansion and volume-acceleration laws~\cite{Buchert2000GERG,BuchertFocus}. 

Defining the volume scale factor as
\begin{equation}
a_{\cD}=\left(V_{\cD}/V_{\cD_i}\right)^{\frac{1}{3}} \ ; \quad V_{\cD_i}=V_{\cD}(t_i) \ ,
\end{equation}
the spatially volume-averaged energy constraint, Raychaudhuri's equation, and mass conservation form a set of exact balance equations that include averaged curvature invariants as backreaction terms:
\bse\label{AppEq:SetAvedEqsQd}
\begin{eqnarray}
3 \left(\frac{\dot{a}_\cD}{a_\cD}\right)^2=\Lambda+ 8 \pi \average{\varrho} - \frac{1}{2}\average{{}^{(3)}\CR}-\frac{1}{2}\CQ_{\cD} \ ; \qquad
\label{AppEq:BackR-da}
\\
3 \frac{\ddot{a}_\cD}{a_\cD}= \Lambda -4 \pi \average{\varrho} + \CQ_{\cD} \ ;  \qquad\label{AppEq:BackR-dda}
\\
\dotaverage{\varrho}= -3 \frac{\dot{a}_\cD}{a_\cD} \average{\varrho} \ ; \qquad\label{AppEq:BackR-rho}
\\
\frac{1}{a_\cD^6}\dotaverage{\CQ_{\cD} a_\cD^6} + \frac{1}{a_\cD^2}\dotaverage{{}^{(3)}\CR a_\cD^2 } =0 \ ,\qquad\label{AppEq:Integrability}
\end{eqnarray}
\ese
where the last equation \eqref{AppEq:Integrability} is redundant in this set;  it follows from the time-derivative of 
\eqref{AppEq:BackR-dda} to yield \eqref{AppEq:BackR-da}.

The rate of change of the volume scale factor, $\dot{a}_\cD/a_\cD$, represents the volume Hubble expansion rate, $H_\cD$, and $\CQ_{\cD}$ is the kinematical backreaction term.
$\CQ_{\cD}$ is expressed through invariants of the expansion tensor (defined as minus the extrinsic curvature of the spatial hypersurfaces), Eq.~\eqref{Eq:QdInv-a}. In the absence of vorticity (our case), it reduces to  Eq.~\eqref{Eq:QdScal-b}. The implementation of this formalism in the LTB subcase of the Szekeres class I solutions was recently examined in~\cite{CliftonSussman2019}.

We have checked that Equations~\eqref{AppEq:SetAvedEqsQd} become identities for the exact functionals of the Szekeres solutions. 

\section{Proof of Lemma~\ref{Lem:lemma2}}\label{App:ProofTheor2}

To proof Lemma~\ref{Lem:lemma2}  in Sec.~\ref{SubSecNoteAdm}, we introduce the rule of noncommutativity (or merely the commutation rule) for the spatial averaging of a scalar field~\cite{Buchert2000GERG},
\begin{equation}\label{AppEq:Commtrule} 
\dotaverage{\Psi}  - \;  \average{\dot \Psi}=\average{\Theta \, \Psi} -\; \average{\Theta} \average{\Psi} \ ,
\end{equation}
where $\Theta$ is the expansion scalar, Eq.~\eqref{Eq:ThetaClassII}.

Let us consider the commutation rule for $\Psi =  P/\mathfrak{J}$ under the assumption $\averageH{\dot P/\mathfrak{J}}=0$,
\begin{eqnarray}
\dotaverageH{\frac{P}{\J}}  &-& \, \averageH{\left(\frac{P}{\J}\right)^{\bdot}}\qquad\nonumber\\
&=&\averageH{\Theta \, \frac{P}{\J}} \!- \,\averageH{\Theta} \averageH{\frac{P}{\J}} \ .
\label{App:EqCommRuleClassII}
\end{eqnarray}
Writing out each term separately,
\begin{subequations}
\begin{eqnarray}
\averageH{\left(\frac{P}{\J}\right)^{\bdot}}= - \averageH{\frac{P \dot P}{\J^2}} \ ,
\\
\averageH{\Theta \, \frac{P}{\J}}= 3 \, \frac{\dot a}{a}\averageH{\frac{P}{\J}} +\, \averageH{\frac{\dot P P}{\J^2}} \ ,
\\
\averageH{\Theta} \averageH{\frac{P}{\J}}= 3 \,\frac{\dot a}{a}\averageH{\frac{P}{\J}} \ ,
\end{eqnarray}
\end{subequations}
and substituting into~\eqref{App:EqCommRuleClassII}, we obtain:
\begin{equation}
\dotaverageH{\frac{P}{\J}}=0 \quad \Rightarrow \quad  \averageH{\frac{P}{\J}}= \text{const.}
\end{equation}
Since the deformation field is initially null, the constant in the above equation is equal to zero.
Hence, we have already proven the first part of the lemma:
\begin{equation}
\averageH{\frac{\dot P}{\J}}=0 \quad \Rightarrow \quad  \averageH{\frac{P}{\J}}=0 \ .
\end{equation}
To proof the second part, let us assume that $\averageH{P / \J}=0$ and examine the commutation rule again. 
The first and last terms of Eq.~\eqref{App:EqCommRuleClassII} are identically zero: the averaged quantity $\averageH{P / \J}$  can be considered a (constant) function of time, $q(t)=0$, so that $\dot{q}(t)=0$. Then:
\begin{equation}
 - \, \averageH{\left(\frac{P}{\J}\right)^{\bdot}} = \averageH{\Theta \, \frac{P}{\J}}  
 \quad \Rightarrow \quad
 \averageH{\frac{\dot P}{\J}} = 0 \ ,
\end{equation}
which proves the second part of the lemma: 
\begin{equation}
 \averageH{\frac{P}{\J}}=0  \quad \Rightarrow \quad   \averageH{\frac{\dot P}{\J}}=0 \ .
\end{equation}
Collecting all results together, we obtain the statement  of Lemma~\ref{Lem:lemma2}:
\begin{equation}
 \averageH{\frac{P}{\J}}=0  \quad \Leftrightarrow \quad   \averageH{\frac{\dot P}{\J}}=0 \ .
\end{equation}
%


\section{Proof of Lemmata~\ref{Lem:ConsRho-3R} and~\ref{Lem:VanishingIntBeta}}\label{AppSec:ProofOfLemmata3-4}

In this section, we will first prove Lemma~\ref{Lem:VanishingIntBeta} and then Lemma~\ref{Lem:ConsRho-3R}, altering the order of their presentation in the main text. 
Let us start by considering the integral constraint~\eqref{Eq:TrueIntConst}, which implies
\begin{eqnarray}
0&=&\Bigl\langle{\frac{P}{\J}}\Bigr\rangle_{\cD_H} = \frac{a^3}{V_{\cD_H}}\left\{\int P \dd^3 \mathbf{X} \right\}
\nonumber \\
&=&-\frac{a^3}{V_{\cD_H}}\left\{\int_{\cD_H} \frac{\F-\F_i}{\A-\F_i} \sqrt{G} \, \dd^3 \rv \right\}
\nonumber \\
&=&-\frac{a^3}{V_{\cD_H}}\left\{\left[\int_{\cD_H} e^{2\nu} \dd \x \dd \y \right]\left[\int_{\cD_H} \left(\F-\F_i\right) \dd \z \right]\right\} \ ,
\nonumber
\\
\end{eqnarray}
where we have used~\eqref{Eq:GabClassIISol},~\eqref{Eq:EclideanIntConst},~\eqref{Eq:P33eqn1}, $\dd^3 {\mathbf X}=\sqrt{G} \,\dd^3 \rv$, and $e^\nu$ was defined in~\eqref{Eq:etonuClassII}. Since $a$, $V_{\cD_H}$, and the first integral in the last line between square brackets are strictly positive, the equality holds, if and only if the second integral vanishes for all $t$, which can be rewritten as follows:
\begin{equation}\label{AppEq:BetaLemProof}
0=q_+(t)\!\int_{\cD_H}\!\!\beta_+( \z) \dd \z + q_-(t)\!\int_{\cD_H}\!\!\beta_-( \z) \dd \z\ .
\end{equation}
In the above equation, we introduced $q_\pm(t)=f_\pm(t)-f_\pm(t_i)$; since they are linearly independent functions, the integral constraint holds for all $t$, if and only if the conditions~\eqref{Eq:TrueIntConstSzeVar} are satisfied. This proves Lemma~\ref{Lem:VanishingIntBeta}.
 
Turning now to the first part of Lemma~\ref{Lem:ConsRho-3R}, and keeping in mind that the density, Eq.~\eqref{Eq:RhoClassII}, can be split into the background term and an exact deviation, we have:
\begin{equation}
\varrho(t,\rv)= \varrho_b(t)\left(1+\frac{\F}{\G}\right) = \varrho_b(t)\left(1+\delta(t,\rv)\right) \ .
\end{equation}
Taking the average, we find that $\averageH{\varrho}= \varrho_b(t)$, if and only if $\averageH{\delta}=\averageH{\F/\G}= 0$, which gives 
\begin{eqnarray}
0&=& \averageH{\delta} 
= \frac{a^3}{V_{\cD_H}} \int  \mathcal{F} e^{2 \nu} d^3 \rv
\nonumber \\
&=& \frac{a^3}{V_{\cD_H}} 
\left[ \int_{\cD_H} e^{2 \nu} \dd \x \dd \y\right] \left[ \int_{\cD_H}  \mathcal{F}\dd \z \right] \ .
\end{eqnarray}
As before, the equality above holds, if and only if  $\int_{\cD_H}  \mathcal{F}\dd \z=0$. Then, writing out this equation, we get an expression similar to~\eqref{AppEq:BetaLemProof}, indicating that~\eqref{Eq:TrueIntConstSzeVar} is a necessary and sufficient condition for the fulfillment of Eq.~\eqref{Eq:MassConClassIILem}.

The second part of the lemma is proved by proceeding along the same lines:   
considering the curvature functional~\eqref{Eq:FuncR3d}, it splits into the FLRW term and an exact deviation, given by 
$-\frac{4}{\J} \lbrack 3\, \frac{ \dot{a} }{ a} \dot{P}+  \ddot{P} \rbrack = \frac{4}{a^2}\lbrack (\beta_+ \left(1+k_0 f_+\right) + k_0 \beta_- f_- ) / ( \A-\F )\rbrack$, see Eq.~\eqref{AppEq:CurvPerturPfs} further below. Then, $\averageH{\CR}=6 k_0/a^2$, if and only if the average of the curvature deviation vanishes. Once again this leads to~\eqref{Eq:TrueIntConstSzeVar} as the necessary and sufficient condition for Eq.~\eqref{Eq:CurvConClassIILem}.
 
\section{Commutation rules for class I }\label{AppSec:ComRulesCIassI}

The spatial dependence of the conformal scale factor makes the commutation rules quite different for class I. Since, in this case, we cannot take the scale factor out of the integral, $P/J$ is not trivially related to $P/\J$.  To illustrate this point, let us examine their commutation rules and some particular cases.     

\subsection*{Commutation rule for $\Psi=P/J$}

Taking $\Psi=P/J$ in~\eqref{AppEq:Commtrule} we obtain:
\begin{equation}
\dotaverage{\frac{P}{J}}  - \average{ \frac{\dot P}{J} } =  -\average{3 \frac{\dot \s}{\s} + \frac{\dot P}{\mathfrak{J}}} \average{\frac{P}{J}}
\ .
\label{Eq:ComRulClasIPdivPJ}
\end{equation}
From this expression, we highlight the following cases:
\begin{itemize}
\item If $\average{P/J}=0$, then: 
\begin{equation}
   \average{ \dot P/J } = 0 \ .
\end{equation}   
\item If $\average{\dot P/J}=0$, then:
\begin{equation}
\dotaverage{\frac{P}{J}}  =  -\average{3 \frac{\dot \s}{\s} + \frac{\dot P}{\mathfrak{J}}} \average{\frac{P}{J}}
\ .
\end{equation}
\end{itemize}

\subsection*{Commutation rule for $\Psi=P/\J$}

Applying the commutation rule to $\Psi=P/\J$ we obtain:
\begin{eqnarray}
\dotaverage{\frac{P}{\mathfrak{J}}}  &-&  \average{\frac{\dot P}{\mathfrak{J}}} 
= 3 \average{ \frac{\dot \s}{\s} \, \frac{P}{\mathfrak{J}}} -
\nonumber
\\
&{}& \;
\left(3\average{ \frac{\dot \s}{\s}}+\average{\frac{\dot P}{\mathfrak{J}}}\right)
 \average{\frac{P}{\mathfrak{J}}} \ .
 \label{Eq:ComRulClasIPdivPfrakJ}
\end{eqnarray}
Similarly to the previous case, we consider the vanishing of $\average{P/\mathfrak{J}}$ and $\average{\dot P/\mathfrak{J}}$:
\begin{itemize}
\item If $\average{\frac{P}{\mathfrak{J}}}=0 $, then:
\begin{equation} \label{Eq:vanishDotPClassI}
    \average{\frac{\dot P}{\mathfrak{J}}}=-3 \average{ \frac{\dot \s}{\s} \, \frac{P}{\mathfrak{J}}} \ .
\end{equation}
\item If $\average{\frac{\dot P}{\mathfrak{J}}}=0 $, then:
\begin{eqnarray}
\dotaverage{\frac{P}{\mathfrak{J}}}
= 3 \average{ \frac{\dot \s}{\s} \, \frac{P}{\mathfrak{J}}} -
3\average{ \frac{\dot \s}{\s}}
 \average{\frac{P}{\mathfrak{J}}} \ .
\end{eqnarray}
\end{itemize}
At first glance, none of these cases provides a useful relation that would render $\CQ_{\cD}=0$ in Eq.~\eqref{Eq:QdClassI}.

\section{Functional evaluation in RZA}\label{SecApp:FuncEvaluationRZA}

This appendix complements Sec.~\ref{SubSec:RZAandCPT} with some formulas and results that, for better readability of the main text, were not included there. Although we will mainly focus on class II, all the results here apply to class I.  

First, we provide the form of Equations~\eqref{Eq:ExactCofraRelations} specialized to our coframe set, needed to obtain the curvature functionals,
\bse
\begin{eqnarray}
J=a^3 \left(1+P\right) \ ;\hspace{9em}
\\
\epsilon_{abc} \epsilon^{mjk} \dot{\eta}^a_{\ m} \dot{\eta}^b_{\ j} \eta^c_{\ k}=2 a\dot a \left[ 3\left(1+P\right)\dot a + 2 a \dot P \right]
\ ; \qquad
 \label{Eq:dn-dn-n}
\\
\epsilon_{abc}\epsilon^{ik\ell} \ddot{\eta}^a_{\ i} \eta^b_{\ k} \eta^c_{\ \ell}=2 a^2 \left[ 2 \dot a \dot P + 3 \left(1+P\right)\ddot a +a \ddot P\right]
\ ; \;
\nonumber
\\
\label{Eq:ddn-dn-n}
\end{eqnarray}
\begin{eqnarray}
 \epsilon_{abc} \epsilon^{ikl} \left( \dot{\eta}^a_{\ j} \eta^b_{\ k} \eta^c_{\ l} \right)^{\bdot} =  \hspace{12em}\nonumber\\
  \begin{cases} 
 2a \Big[ 
2 \left(1+P\right) \dot{a}^2+a \dot{a} \dot{P} +a \left(1+P\right)  \ddot{a} \Big] , \;  i=j=1, 2 \ , 
 \\ 
 2a \Big[ 
 2 \left(1+P\right) \dot{a}^2+4a \dot{a} \dot{P} \nonumber\\ \hspace{7em} +\ a  \left( \left(1+P\right) \ddot{a} +a \ddot{P}\right)
 \Big], \;  i=j=3 \ , 
 \\
0 \ , \ \hbox{otherwise ;}  
\end{cases}
\nonumber
\\
\label{Eq:d_dn-dn-n_ij}
\end{eqnarray}
\begin{eqnarray}
\t{\epsilon}{-}{abc} \t{\epsilon}{ikl}{} \addot{a}{j} \a{b}{k} \a{c}{l}=
 \hspace{12em}
\nonumber\\
\begin{cases} 
2 a^2 \ddot a  \left(1+P\right) , \hspace{7em}  i=j=1, 2 \ , 
 \\ 
2 a^2 
\left[
2 \dot a \dot P +\left(1+P\right) \ddot a + a \ddot P
\right]
 \ , \  i=j=3 \ , 
 \\
0 \ ,  \ \hbox{otherwise .}
\end{cases}
\label{Eq:d2n-n-n}
\end{eqnarray}
\ese
In what follows, we will prove that all the functionals provided in the main text, reduce to the Szekeres quantities when we substitute the expression for the deformation field in terms of the Szekeres functions.

Let us start by writing out the functionals for the peculiar-determinant and $\dot P/\J$,  
\begin{eqnarray}
\J&=&1+ P = 1-\frac{\F-\F_{i}}{\A-\F_i}=\frac{\A-\F}{\A-\F_i} \ ;
\\
\frac{\dot P}{\J}&=&\left(-\frac{\dot\F}{\A-\F_i}\right)\left(\frac{\A-\F}{\A-\F_i}\right)^{-1} 
=-\frac{\beta_+ \dot{f}_+ + \beta_- \dot{f}_-}{\G} \ .
\nonumber
\\
\end{eqnarray}
Then, our functionals reproduce the correct results for the energy density, 
\begin{equation}\label{EqApp:RhoFunc}
4\pi \varrho=4\pi \frac{ \varrho(t_i)}{J^{-1}}=4 \pi \frac{  \varrho_b(t_i)}{a^3}\left(1+\frac{\F}{\G}\right) \ ,
\end{equation}
the expansion scalar and the shear tensor (see (2.19)-(2.20) in~\cite{GW1}), 
\bse
\begin{eqnarray}
\Theta&=&3 \, \frac{\dot{a}}{a}  +\frac{ \dot P }{\J}=3 \, \frac{\dot{a}}{a} - \frac{\beta_+ \dot{f}_+ + \beta_- \dot{f}_-}{\A-\F} \ ;
\\
2\t{\sigma}{1}{1}&=&2\t{\sigma}{2}{2}=-\t{\sigma}{3}{3}=2\Sigma = -\frac{2}{3} \, \frac{\dot P}{\J}  
\nonumber \\
&{}&\hspace{5.8em}=\frac{2}{3} \,\frac{\beta_+ \dot{f}_+ + \beta_- \dot{f}_-}{\A-\F} \ . 
\end{eqnarray}
\ese
In Equation~\eqref{EqApp:RhoFunc} above, $3\mu=4\pi \varrho_b(t_i)$ and  $\varrho(t_i)= \varrho_b(t_i)\left(1+ \F_i / \G_i \right)$.

The only nontrivial  Weyl scalar,  $\Psi_2$, is rewritten as
\begin{equation}
\Psi_2  = \frac{1}{  3\, \J}
\Bigg \{
   2 \frac{\dot a}{a} \dot P + \ddot P
\Bigg \}
 = -\frac{\mu}{  a^3 } 
\Bigg \{
\frac{\beta_+ f_+ + \beta_- f_- }{\G}
\Bigg \} \ , 
\label{AppEq:Psi2Sze}
\end{equation}
where we have used Equation~\eqref{Eq:Ftt} to get~\eqref{AppEq:Psi2Sze}. The final gravitoelectric part of the Weyl tensor reads (which is the same as Equation (2.21) in~\cite{GW1}):
\begin{equation}
2\t{E}{1}{1}=2\t{E}{2}{2}=-\t{E}{3}{3}=-\frac{2 \mu}{  a^3 } 
\Bigg \{
\frac{\beta_+ f_+ + \beta_- f_- }{\G}
\Bigg \} \ .
\end{equation}
To obtain the expressions for the curvature, let us first rewrite the following quantity in terms of the Szekeres functions,
\begin{align}
{}&
\frac{1}{\J}\Big(3 \, \frac{\dot a}{a}\dot{P}+\ddot P\Big) =
\nonumber
\\
{}&\quad -\frac{ \beta_+ \left(\ddot{f}_+ + 3\, \frac{\dot a}{a} \dot{f}_+ \right) + \beta_- \left(\ddot{f}_- + 3\, \frac{\dot a}{a} \dot{f}_- \right)}{\G} \ . 
\label{AppSEq:ExprforRandT}
\end{align}
This equation coincides with the one obtained for the  components of the trace-free spatial Ricci tensor in the original Goode-Wainwright paper---see Appendix B of~\cite{GW1}, taking into account that $\t{\tau}{3}{3}=-\frac{2}{3\J}
\left(3 \, \frac{\dot a}{a}\dot{P}+\ddot P\right)$. To  work it out towards a simpler expression, without time-derivatives, we need to introduce the first integral of~\eqref{Eq:Ftt}, Eq.~\eqref{Eq:1st-IntFtt}, rewritten below as follows:
\begin{subequations}
\label{AppEq:1stInteFtt}
\begin{eqnarray}
a \,\dot{a} \, \dot{f}_+ -\left(k_0-\frac{3\mu}{a}\right)\, f_+=1 \ ;
\\ 
a \,\dot{a} \, \dot{f}_- -\left(k_0-\frac{3\mu}{a}\right)\, f_-=0 \ .
\end{eqnarray}
\end{subequations}
Using~\eqref{AppEq:1stInteFtt}, the terms in parentheses in~\eqref{AppSEq:ExprforRandT} can be rewritten as
\begin{equation}
\ddot{f}_\pm + \,3 \, \frac{\dot a}{a} \dot{f}_\pm = \ddot{f}_\pm + 2 \, \frac{\dot a}{a} \dot{f}_\pm + \frac{\dot a}{a} \dot{f}_\pm= \left(k_0 f_\pm + \alpha_\pm \right) a^{-2} \ ,
\end{equation}
where $\alpha_+ =1$ and $\alpha_- =0$. Then,
\begin{equation}\label{AppEq:CurvPerturPfs}
\frac{1}{\J}
\Big(3 \, \frac{\dot a}{a}\dot{P}+\ddot P\Big) 
= -\frac{
\beta_+ \left(1+k_0 f_+\right) + k_0 \beta_- f_- 
}{a^2 \G} \ .
\end{equation}
Substituting this expression into~\eqref{Eq:FuncR3d} and~\eqref{Seq:TauFuncEval}, we arrive at the final expressions for the spatial  Ricci scalar and the trace-free Ricci tensor,
\begin{eqnarray}
\CR&=&
\frac{6}{a^2}
\left\{
k_0+\frac{2}{3\G}\left[\beta_+ \left(1+k_0 f_+\right) + k_0 \beta_- f_-\right]
\right\}; 
\nonumber
\\
\\
2\t{\tau}{1}{1} &=&2\t{\tau}{2}{2}=-\t{\tau}{3}{3}
\nonumber
\\
&=&-\frac{2}{3 a^2 \G} 
\left[ \beta_+ \left(1+k_0 f_+\right) + k_0 \beta_- f_-  \right] \ .
\end{eqnarray}
%


\section{LTB subcase}\label{App:LTBSubcase}

The LTB solution emerges from the Szekeres-Szafron parametrization, as a subcase of the quasispherical models of class I, when the free functions $A(\z)$,  $B_1(\z)$, $B_2(\z)$ and $C(\z)$ are all constants ($\nu_{,\z}=0$ in~\eqref{nudefeq1}), the coordinate $\z$ is identified with the radial comoving coordinate $r$, $k\rightarrow-2 E$,  $\Phi\rightarrow R$, and $(\x,\y)$ are transformed to the angular coordinates $(\vartheta,\varphi)$ by a stereographic transformation---see~\cite{SussBol2012, hellaby1996null} for suitable parametrizations from where the LTB solution arises in more straightforward ways.

Instead of obtaining the spherically symmetric limit directly from the GW formulation, we can identify the LTB deformation field by tracing back the metric coefficients from one parametrization to the other. 
If we factor out the function $\Phi$ in the metric coefficient $g_{\z \z}$ of the Szekeres-Szafron parametrization, then we can associate $\A$ (in the GW parametrization) with $\nu_{,\z}$ (in the Szekeres-Szafron parametrization). Since they both vanish in the LTB limit, the contribution of $\A$  will either vanish or be absorbed by the scale function reparametrization. 
The remaining terms will contribute to the Gram's matrix or be wiped out by the stereographic transformation. Thus, $\F$ will be associated to the quotient of the scale factor and its derivative: $\F\rightarrow R^{\prime}/R$, see Eq.~\eqref{EqApp:DefFeq1}. 

The above reasoning is supported by what we get directly from the LTB line-element:
\begin{eqnarray}
\dd s^2 &=& -\dd t^2 + \frac{R^{\prime 2}}{1+2 E} \dd r^2  + R^2 \dd \Omega^2 \nonumber
\\
&=&-\dd t^2 + R^2 \left( \frac{\Gamma^2}{1+2 E} \dd r^2  + \dd \Omega^2\right) \ ,
\end{eqnarray}
with $\Gamma\equiv R^\prime/R$.

\subsection{LTB solution in the language of RZA}

The deformation field and Gram's matrix associated with the LTB solution are given by: 
\begin{equation}\label{SEqs:P_ltb}
P=\t{P}{3}{3}=-\frac{\widetilde{\F}}{\widetilde{\A}}\equiv-\frac{\F-\F_i}{\A-\F_i}\rightarrow \frac{\Gamma-\Gamma_i}{\Gamma_i} \ ,
\end{equation}
and
\begin{equation}
G_{ab}=\hbox{Diag}\left[\frac{\Gamma_i^{~2}}{1+2E}\ ,\, \sin^2 \vartheta \ , \, 1 \right] \ .
\end{equation}
Substituting~\eqref{SEqs:P_ltb} into the  ``exact body functionals'' for expansion scalar and shear tensor, and taking into account that
\begin{equation}
\frac{\dot P}{\J}=\left(\frac{\dot{R}^\prime}{R^\prime}-\frac{\dot R}{R}\right) 
\;
\Rightarrow
\;
\Sigma=-\frac{1}{3}\frac{\dot P}{\J}=\frac{1}{3} \left(\frac{\dot R}{R}-\frac{\dot{R}^\prime}{R^\prime}\right) 
\ ,
\end{equation}
we obtain the correct results for the LTB scalar expansion rate and shear tensor:
\bse
\begin{align}
\Theta&=2\frac{\dot R}{R} + \frac{\dot{R}^\prime}{R^\prime} \ ;
\\
2\t{\sigma}{\vartheta}{\vartheta}&=2\t{\sigma}{\varphi}{\varphi}=-\t{\sigma}{r}{r}=\frac{2}{3} \left(\frac{\dot R}{R}-\frac{\dot{R}^\prime}{R^\prime}\right) \ .
\end{align}
\ese
The volume element takes its well-known form:
\begin{eqnarray}
J \, \dd^3 \mathbf{X}&=&R^3 \left(1+P\right) \sqrt{G} \, \dd^3 \mathbf{x}  
\nonumber
\\
&=&
R^3 \frac{\Gamma}{\Gamma_i} \frac{\Gamma_i \sin \vartheta}{(1+2 E)^{1/2}}\dd^3 \mathbf{x} \nonumber
\\
&=&
R^2 R^\prime  \frac{\sin \vartheta}{(1+2 E)^{1/2}}\dd^3 \mathbf{x} \ .
\end{eqnarray}

\subsection{Some general features of the LTB models}

The integration of the Einstein equations reduces to a nonlinear ordinary differential equation that resembles the equation of movement in a Newtonian Coulomb potential (for a comprehensive and detailed exposition of the LTB solution see Section $18$ of~\cite{kras2} and Section $2.1$ of~\cite{BKHC2009}). Including the cosmological constant, we have
\begin{equation}\label{AppEq:LTBEq-a}
\dot{R}^2=2 E(r) + \frac{2 M(r)}{R} +\frac{\Lambda R^2}{3}  \ ,
\end{equation}
where the function $M(r)$ is arbitrary. 

The model is characterized by three free parameters: $E(r)$, $M(r)$ and $t_b(r)$. The latter, the ``big bang time'', arises from the integration of~\eqref{AppEq:LTBEq-a},
\begin{eqnarray}
t-t_b(r)=\int_0^R \frac{\dd \hat{R}}{\sqrt{2 E(r) + \frac{2 M(r)}{\hat{R}} +\frac{\Lambda \hat{R}^2}{3} }} \ .
\end{eqnarray}
In order to have a regular solution, the arbitrary free functions should satisfy the following regularity conditions at the symmetry center (origin of coordinates): 
\begin{eqnarray}
R(t,0)=0  \quad \wedge  \quad M(0)=0 \ , \quad \text{with} \quad
\\
R\propto M^{1/3}  \quad \wedge \quad E \propto M^{2/3}  \quad \text{at} \quad r\rightarrow 0 \ .
\end{eqnarray}
Additionally, the gradients of $M(r)$, $E(r)$ and $t_b(r)$ identically vanish at $r=0$.

In these coordinates, the FLRW solution results from the limit,
\begin{equation}
M=M_0 r^3 \ , \quad E=- \frac{1}{2} k_0 r^2 \ , \quad t_b=\text{const.\ ,} 
\end{equation}
where $M_0$ and $k_0$ denote arbitrary constants.

\subsection{Averaging and backreaction in LTB models}

We can reconstruct the  expansion tensor from the above expressions for the shear and scalar expansion, $\t{\Theta}{i}{j}=\t{\sigma}{i}{j}+\frac{1}{3}  \Theta \,\t{\delta}{i}{j}$. Then, its principal scalar invariants read:
\bse
\begin{eqnarray}
\inI(\Theta_{\; j}^{i})&=&2\frac{\dot R}{R} + \frac{\dot{R}^{\prime}}{R^{\prime}}\ ;
    \\
\inII(\Theta_{\; j}^{i})&=& \frac{\dot{R}^2}{R^2}+2 \frac{\dot{R} \dot{R}^{\prime}}{R R^\prime}
\ ;
    \\
\inIII(\Theta_{\; j}^{i})&=& \frac{\dot{R}^2 \dot{R}^\prime}{R^2 R^\prime} \ .
\end{eqnarray}
\ese
Averaging over a spherical, compact domain $\cB$ we obtain:
\bse\label{AppSubEq:GenIntInvLTB}
\begin{eqnarray}
\averageL{\inI}=\frac{4 \pi}{V_{\cB}} \int_0^{r_{\cB}}\frac{\partial_r \left(\dot{R}R^2\right)}{\sqrt{1+2 E}} \dd r \ ;
\\
\averageL{\inII}=\frac{4 \pi}{V_{\cB}} \int_0^{r_{\cB}}\frac{\partial_r \left(\dot{R}^2 R\right)}{\sqrt{1+2 E}} \dd r \ ;
\\
\averageL{\inIII}=\frac{4 \pi}{3 V_{\cB}} \int_0^{r_{\cB}}\frac{\partial_r \left(\dot{R}^3\right) }{\sqrt{1+2 E}}\dd r \ ,
\end{eqnarray}
where
\begin{equation}
V_{\cB}= 4 \pi \int_0^{r_{\cB}} \frac{R^2 R^{\prime}}{\sqrt{1+2 E}} \dd r \ .
\end{equation}
\ese
In general, this set of integrals  cannot be evaluated explicitly except for some special cases. Below, we discuss two important specializations of the arbitrary functions that lead to the vanishing of the backreaction term on $\cB$.

\subsubsection*{Case $E=0$ (parabolic models)} 

For parabolic solutions,  the integrals~\eqref{AppSubEq:GenIntInvLTB} yield:
\bse
\begin{eqnarray}
V_{\cB}&=& \frac{4 \pi}{3} R^3_{\cB} \ ,  \qquad \qquad {\rm and}\ ,
\\
\averageL{\inI}&=&3 \frac{\dot{R}_{\cB}}{R_{\cB}} \ ; \;\; \averageL{\inII}=3 \left(\frac{\dot{R}_{\cB}}{R_{\cB}}\right)^2 \ ; \;\; \averageL{\inIII}= \left(\frac{\dot{R}_{\cB}}{R_{\cB}}\right)^3 \ , 
\nonumber
\\
\end{eqnarray}
\ese
with $R_{\cB}\equiv R(t,r_{\cB})$. These relations result in the vanishing of the kinematical backreaction on $\cB$:
\begin{equation}\label{AppEq:QB0Inv}
\CQ_{\cB}= 2\averageL{\inII} - \frac{2}{3} \averageL{\inI}^2=0 \ .
\end{equation}

\subsubsection*{Case $R=\psi(t) \cdot \chi(r)$ (type-class-II or separable solution) } 

The separable solutions of the LTB models, with
\begin{equation}\label{AppEq:LTBSeparable}
R=\psi(t) \cdot \chi(r) \ ,
\end{equation}
correspond to another case with no backreaction (on an arbitrary compact domain $\cB$). However, what is somewhat surprising is that this solution is nothing more than the FLRW solution in a nonstandard coordinate system. 
To verify this statement, let us note that the separable ansatz  corresponds to a shear-free model.
Substituting~\eqref{AppEq:LTBSeparable} into the expression for the degenerate eigenvalue of the shear tensor, we obtain:
\begin{equation}
\Sigma=\frac{1}{3} \left(\frac{\dot R}{R}-\frac{\dot{R}^\prime}{R^\prime}\right) 
= \frac{1}{3} \left(\frac{\dot{\psi} \,  \chi }{\psi \,  \chi}-\frac{\dot{\psi } \, \chi^\prime}{\psi \, \chi^\prime}\right)
=0 \ .
\end{equation}
Its identification as the FLRW model directly follows from  a coordinate-independent  definition of this class (Pleba\'nski and Krasi\'nski~ \cite{kras2}); the necessary  and sufficient conditions for a solution to be included in the FLRW class can be enunciated as follows: 
\begin{itemize}
\item[(i)] It is an exact solution of the Einstein equations with a perfect fluid source.
\item[(ii)] The fluid flow has neither rotation, acceleration, nor shear.
\end{itemize}
For the general LTB model, the vanishing of the shear tensor is the only one of these conditions that is not identically satisfied. However, in our case, $\t{\sigma}{a}{b}=0$ is a consequence of the additional separability assumption, which renders the solution a FLRW model. 

\hfill

Overall, the vanishing of $\CQ_{\cB}$ is only true for the parabolic case that corresponds to the Newtonian ``iron sphere theorem''.
For LTB models with curvature, as well as for Szekeres class I models, a closed-form expression for 
the kinematical backreaction functional, {\it e.g.} in terms of the volume, is an open question. 

For some further discussions of average properties of LTB solutions in relation to RZA, see [Sect.7]\cite{BuchertFocus}, and of silent universe models, see [Sect. 6.2]\cite{GBC}.

\subsection{More on the evaluation of $\CQ_{\cB}$ for parabolic LTB models}

In this appendix, we aim at complementing the discussion in Sect.~\ref{SubSec:AdmissibDataClassI} by showing that for  parabolic LTB models, the vanishing of the kinematical backreaction does not reveal any trivial relation between the variables in the expression~\eqref{Eq:QdClassI} for $\CQ_{\cB}$. 

Averaging the parabolic solution on $\cB$, we obtain the following relations:
\bse\label{AppEq:TermsQB-R}
\begin{eqnarray}
\averageL{\left(\frac{\dot R}{R}\right)^2} &\ =\ &  \frac{3}{R^3_{\cB}} \int_0^{r_{\cB}} \dot{R}^2 R^\prime \dd r
=\frac{3}{R^3_{\cB}} I_1 \ ;\quad
\\
\averageL{\frac{\dot R}{R}}&\ =\ & \frac{3}{R^3_{\cB}} \int_0^{r_{\cB}} \dot{R} R^\prime R \, \dd r
=\frac{3}{R^3_{\cB}} I_2 \ ;\qquad\qquad
\end{eqnarray}
\begin{eqnarray}
 \averageL{\Sigma}&\ =\ & \frac{1}{R^3_{\cB}} \left( \int_0^{r_{\cB}} \dot{R} R^\prime R \, \dd r -\int_0^{r_{\cB}} \dot{R}^\prime R^2 \, \dd r \right) 
\nonumber
\\
 &=&\frac{1}{R^3_{\cB}} \left( I_2 -  I_3 \right) \ ;
\\
 \averageL{\frac{\dot R}{R}\Sigma}\!\! \!&=&\!\! \frac{1}{R^3_{\cB}} \left( \int_0^{r_{\cB}} \dot{R}^2 R^\prime \, \dd r -\int_0^{r_{\cB}}  \dot{R} \dot{R}^\prime R \, \dd r \right) 
 \nonumber
\\
&=&\frac{1}{R^3_{\cB}} \left( I_1 -  I_4 \right) \ ,
\end{eqnarray}
\ese
where, in general, neither of these terms vanish. In the above equations we have defined:
\bse
\begin{eqnarray}
I_1=\int_0^{r_{\cB}} \dot{R}^2 R^\prime \dd r  \ ; \quad I_2=\int_0^{r_{\cB}} \dot{R} R^\prime R \, \dd r \ ;\quad
\\
I_3= \int_0^{r_{\cB}} \dot{R}^\prime R^2 \, \dd r  \ ; \quad I_4=\int_0^{r_{\cB}}  \dot{R} \dot{R}^\prime R \, \dd r\ .\quad
\end{eqnarray}
\ese
To make clear that these equations reduce to~\eqref{AppEq:QB0Inv}, we have to manipulate some of the integrals above. Integrating by parts leads to the following relations:
\begin{eqnarray}
I_1=\dot{R}^2_{\cB} R_{\cB}-2 I_4 \ ; \quad I_3=\dot{R}_{\cB} R^2_{\cB}-2 I_2 \ , 
\end{eqnarray}
which yields for the backreaction term:
\begin{eqnarray}
\frac{1}{6}\CQ_{\cB}
&=&
\averageL{\left(\frac{\dot R}{R}\right)^2} 
- \averageL{\frac{\dot R}{R}}^2 - \averageL{\Sigma}^2
\qquad\quad
\nonumber
\\
&{}&
+\ 2 \averageL{\frac{\dot R}{R}}\averageL{\Sigma}- 2 \averageL{\frac{\dot R}{R}\Sigma}
\nonumber
\\
&=&
\frac{\left(I_1+2 I_4\right) R^3-\left(I_3+2 I_2\right)^2}{R^6}
= 0 \ .
\end{eqnarray}
Let us calculate each of the terms above  
for the case with a vanishing cosmological constant, where the analytical solution reads:   
\begin{equation}
R(r,t)=\left[\frac{9}{2}M(r)\left(t-t_{b}(r)\right)^2\right]^{1/3} \ .
\end{equation}
Carrying out the integrals in~\eqref{AppEq:TermsQB-R},
\bse
\begin{eqnarray}
\averageL{\left(\frac{\dot R}{R}\right)^2} =
\frac{4}{9}\frac{M-2 \int_0^{r_{\cB}} \frac{ M t_b^\prime}{t-t_b} \dd \chi}{ M \left(t-t_b\right)^2} 
=\frac{4}{9}\frac{M-2 \mathcal{I}_1}{ M \left(t-t_b\right)^2} \ ; \nonumber\\
\end{eqnarray}
\begin{eqnarray}
\averageL{\frac{\dot R}{R}} =
\frac{2}{3}\frac{\int_0^{r_{\cB}} \left(t-t_b\right) M^\prime\dd \chi - 2\int_0^{r_{\cB}} M t_b^\prime \dd \chi}{ M \left(t-t_b \right)^2} 
\nonumber\\
=\frac{2}{3}\frac{\mathcal{I}_2 - 2 \mathcal{I}_3}{ M \left(t-t_b \right)^2} \ ;\qquad
\end{eqnarray}
\begin{eqnarray}
\averageL{\frac{\dot R}{R}\Sigma} = -\frac{4}{9}\frac{\int_0^{r_{\cB}} \frac{M t_b^\prime}{t-t_b} \dd \chi}{ M \left(t-t_b \right)^2} = -\frac{4}{9}\frac{\mathcal{I}_1}{ M \left(t-t_b \right)^2}  \ ;\nonumber\\
\end{eqnarray}
\begin{eqnarray}
 \averageL{\Sigma} &=& -\frac{2}{3}\frac{\int_0^{r_{\cB}} M t_b^\prime \dd \chi}{M \left(t-t_b \right)^2}
=-\frac{2}{3}\frac{\mathcal{I}_3}{M \left(t-t_b \right)^2} \ ,\quad
\end{eqnarray}
\ese
where: 
\bse
\begin{eqnarray}
\mathcal{I}_1&=&\int_0^{r_{\cB}} \frac{ M t_b^\prime}{t-t_b} \dd \chi \ ;
\; \mathcal{I}_2 =\int_0^{r_{\cB}} \left(t-t_b\right) M^\prime\dd \chi \ ;
\qquad\;\;
\\
\mathcal{I}_3&=&\int_0^{r_{\cB}} M t_b^\prime \dd \chi \ ; 
\;
V_{\cB}=6\pi M(r)\left(t-t_b(r)\right)^2 . \qquad\;\;\;
\label{volumeLTB}
\end{eqnarray}
\ese
For the backreaction term we then obtain:
\begin{equation}
\frac{1}{6}\CQ_{\cB}
=\frac{4}{9}
\left(
\frac{1}{ \left(t-t_b\right)^2}
-\frac{\left({\mathcal I}_2-{\mathcal I}_3\right)^2}{M^2 \left(t-t_b \right)^4}
\right) \ ,
\end{equation}
where the numerator of the last term in parentheses simplifies to
\begin{eqnarray}
\mathcal{I}_2-\mathcal{I}_3&=&\int_0^{r_{\cB}} \left(t-t_b\right) M^\prime\dd \chi - \int_0^{r_{\cB}} M t_b^\prime \dd \chi 
\nonumber
\\
&=&\int_0^{r_{\cB}} \left(\left(t-t_b\right) M\right)^\prime\dd \chi 
= M \left(t-t_b\right) \ . \qquad
\end{eqnarray}
Hence,
\begin{eqnarray}
\frac{1}{6}\CQ_{\cB}
&=&\frac{4}{9}
\left(
\frac{1}{ \left(t-t_b\right)^2}
-\frac{\left(M \left(t-t_b\right)\right)^2}{M^2 \left(t-t_b \right)^4} \ \right) 
\nonumber
\\
&=&
\frac{4}{9}
\left(
\frac{1}{ \left(t-t_b\right)^2}
-\frac{1}{ \left(t-t_b \right)^2}
\right)
=0 \ .
\end{eqnarray}\\
Alternatively, we can arrive at the same result directly from the invariants of the expansion tensor. Carrying out~\eqref{AppSubEq:GenIntInvLTB}, we obtain:
\bse
\begin{eqnarray}
&\averageL{\inI} =\frac{2}{t-t_b} \ ; \\
&\averageL{\inII} =\frac{1}{3}\frac{4}{\left(t-t_b\right)^2}=\frac{1}{3} \averageL{\inI}^2 \ ; 
\\
&\averageL{\inIII} = \frac{8}{27 (t-t_b)^3} = \frac{1}{27}\averageL{\inI}^3 \ .
\end{eqnarray}
\ese
These relations lead to the same result for the backreaction term, $\CQ_{\cB}= 2\averageL{\inII} - \frac{2}{3} \averageL{\inI}^2=0$.
\end{appendix}
\newcommand\eprintarXiv[1]{\href{http://arXiv.org/abs/#1}{arXiv:#1}}

\end{document}